# Isoenergetic Two-Photon Excitation Enhances Solvent-to-Solute Excited-State Proton Transfer

Jurick Lahiri[1], Mehdi Moemeni[1], Jessica Kline[1], Ilias Magoulas[1], Stephen H. Yuwono[1], Maryann Laboe[2], Jun Shen[1], Babak Borhan[1,*], Piotr Piecuch[1,3,*], James E. Jackson[1,*], G. J. Blanchard[1,*], and Marcos Dantus[1,3,*#]

[1] Department of Chemistry, Michigan State University, East Lansing, MI 48824, USA
[2] Department of Chemical Engineering and Material Science, Michigan State University, East Lansing, MI 48824, USA
[3] Department of Physics and Astronomy, Michigan State University, East Lansing, MI 48824, USA

* Corresponding authors: BB email: babak@chemistry.msu.edu, tel.: +1-517-353-0501; PP email: piecuch@chemistry.msu.edu, tel.: +1-517-353-1151; JEJ email: jackson@chemistry.msu.edu, tel.: +1-517-353-0504; GJB email: blanchard@chemistry.msu.edu, tel.: +1-517-353-1105; MD email: dantus@chemistry.msu.edu, tel.: +1-517-353-1191. [#] Lead contact.

**ABSTRACT**

Two-photon excitation is an attractive means for controlling chemistry in both space and time. Isoenergetic one- and two-photon excitations (OPE and TPE) in non-centrosymmetric molecules are often assumed to reach the same excited state and, hence, to produce similar excited-state reactivity. We compare the solvent-to-solute excited-state proton transfer of the super photobase **FR0**-SB following isoenergetic OPE and TPE. We find up to 62 % increased reactivity following TPE compared to OPE. From steady-state spectroscopy, we rule out the involvement of different excited states and find that OPE and TPE spectra are identical in non-polar solvents but not in polar ones. We propose that differences in the matrix elements that contribute to the two-photon absorption cross sections lead to the observed enhanced isoenergetic reactivity, consistent with the predictions of our high-level coupled–cluster-based computational protocol. We find that polar solvent configurations favor greater dipole moment change between ground and excited states, which enters the probability for two-photon excitations as the absolute value squared. This, in turn, causes a difference in the Franck-Condon region reached via TPE compared to OPE. We conclude that a new method has been found for controlling chemical reactivity via the matrix elements that affect two-photon cross sections, which may be of great utility for spatial and temporal precision chemistry.

# I. INTRODUCTION

Two-photon excitation[1] (TPE) is an attractive means of chemical activation because it allows one to control chemical processes in space and time with resolution limited only by the laser pulse used, typically sub-micron spatial resolution and sub-picosecond temporal resolution. The high spatial resolution achieved via TPE led to the development of multi-photon microscopy, which is capable of providing sub-micron resolution through scattering biological tissues.[2–7] These advantages are particularly important when imaging strongly absorbing samples, such as blood, or highly sensitive tissues, *e.g.*, the retina.[8,9] Similarly, TPE has been adopted as a valuable method for sub-micron photolithography.[10–14] As part of an effort to develop the tools required for precision chemistry, where chemical reactions can be activated and deactivated with high temporal and spatial control, we have evaluated if strong photobases[15] can be made better photoreagents through the use of TPE.

Precision chemistry of light-induced acid–base reactions requires controlling the underlying excited-state proton transfer (ESPT) processes.[16–20] This broad category of chemical reactions can generally be divided into reversible and irreversible and intramolecular and intermolecular. Here, we focus on reversible intermolecular processes that may be amenable to precision chemistry. From the point of view of the photo-activated reagent, there are numerous proton-donating species, called photoacids, essentially hydroxylated aromatic compounds, while proton-abstracting molecules, *i.e.*, photobases, are less common. The present work examines the super photobase **FR0**-SB (7-((butylimino)methyl)-*N*,*N*-diethyl-9,9-dimethyl-9*H*-fluoren-2-amine), a non-centrosymmetric fluorene Schiff base shown in Fig. 1(a), capable of abstracting protons from alcohols ranging from methanol to *n*-octanol.[15,21] While other compounds, primarily quinoline derivatives, have been found to undergo ESPT in methanol, with 5-methoxyquinoline reaching an excited-state $pK_a$ value of 15.5,[22–29] our work has focused on **FR0**-SB because of its stronger photobasicity ($pK_a^* = 21$).[15,21] Reversible photobases are relatively scarce, because their reactivity depends on having a high excited-state $pK_a$ and the ability to abstract a proton from the solvent within the lifetime of the excitation and the relevant solvent reorganization time. In particular, solvation of the resulting ions has been found to require two or more solvent molecules in a specific configuration.[22,27,30–37]

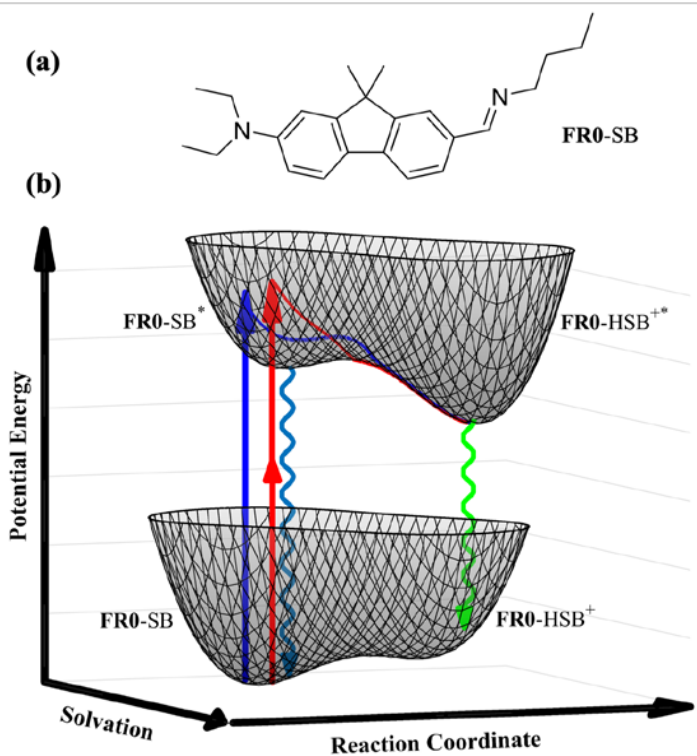

**FIG. 1**. (a) Structural formula of **FR0**-SB. (b) Schematic representation of the ground- and excited-state potential energy surfaces of **FR0**-SB along the solvation and ESPT reaction coordinates. We illustrate how OPE (blue) and TPE (red) to the first excited singlet $S_1$ state of **FR0**-SB may favor differently solvated molecules from the inhomogeneous ensemble and result in accessing different Franck-Condon regions. Note that OPE and TPE are accessing the same excited electronic state. Different entries into the $S_1$ potential energy surface lead to different reaction trajectories toward protonation, with different dynamics and different probability to reach the protonated excited state. Fluorescence from both the non-protonated (blue wiggly line) and protonated (green wiggly line) excited states provide information on the progress of the reaction.

The primary focus of this work is to explore the reactivity of **FR0**-SB upon two-photon photo-activation. **FR0**-SB lacks a center of inversion, so Laporte's symmetry rule preventing one- and two-photon transitions to the same excited state does not apply. Therefore, both one-photon excitation (OPE) and TPE to the first excited singlet ($S_1$) state of **FR0**-SB are allowed. While the excitation efficiencies for OPE and TPE are quite different, one might reasonably expect similar reactivity following isoenergetic (in the case of this work, $\omega_{OPE} = 2\ \omega_{TPE}$) excitation. For **FR0**-SB in alcohol solvents, however, we find the extent of solvent-to-solute ESPT following TPE to be as much as 62 % greater than that following OPE. We present direct evidence for this surprising finding through steady-state and time-resolved spectroscopic data. We discuss several hypotheses and support or refute them based on experimental findings and theoretical calculations. Finally, we conclude that the molecular properties governing TPE, which we estimate from the spectroscopic data as well as using high-level quantum chemistry computations, lead to the formation of an excited-state wave packet at a different Franck-Condon region compared to OPE, thus changing the entry point onto the excited-state potential energy surface and, consequently, giving rise to a different trajectory along the reaction coordinate (see Fig. 1).

## II. EXPERIMENTAL METHODS

For the OPE and TPE fluorescence measurements, tunable ~50 fs pulses centered at 800 nm were obtained from a noncolinear optical parametric amplifier (Orpheus-N-3H, Light Conversion) pumped by the third-harmonic of a Pharos Yb:KGW laser producing 50 mJ of pulse energy at a repetition rate of 200 kHz centered at 1030 nm. For OPE, the output was frequency doubled by a β-BBO crystal to produce excitation light centered at 400 nm focused by a 10 cm focal length convex lens onto a 1 cm cuvette. The sample solutions had a ~3 μM concentration, corresponding to an optical density of less than 0.2. For TPE, the output was focused by a 20 cm focal length convex lens to the same sample. The fluorescence was captured with an optical fiber with detection using an Ocean Optics QE PRO spectrometer for measurements in methanol and ethanol, while an Ocean Optics QE65000 instrument was used in the case of *n*-propanol, *i*-propanol, and *n*-hexanol.

For time-resolved fluorescence measurements, a Ti:Sapphire oscillator (Coherent Vitara-S) producing pulses at 80 MHz centered at 800 nm was used for laser excitation. For OPE, the frequency doubling of the laser output was achieved with a β-BBO crystal and the sample was excited with linearly polarized (vertical) laser pulses. For TPE, the laser pulses were polarization-rotated by 90° with a half-wave plate to maintain the same linear (vertical) polarization between TPE and OPE excitation. The **FR0**-SB sample with an optical density of 0.2 or below was contained in a 1 cm cuvette. Fluorescence emitted at right angles was acquired at parallel and perpendicular polarizations with respect to the vertically polarized excitation pulse using a polarizer followed by detection with a 16-multiplier time-correlated single-photon counting (TCSPC) system (SPC-830 TCSPC, Becker-Hickl, GmBH). The reported time constants were obtained after extracting the isotropic component from the fluorescence decays and fitting with a convolute-and-compare routine to account for the instrument response function. Details of the fitting procedure can be found in the supplementary material. Measurements were repeated at least 5 times for each solvent to quantify uncertainties.

The TPE spectra of **FR0**-SB in methanol, acetonitrile, and cyclohexane were measured in the same optical setup as the TPE fluorescence, except for a glass slide, which was introduced in the path of the excitation beam before the converging lens, to reflect part of the beam to be scattered on a diffuser and detected with a compact Ocean Optics spectrometer. The scattered integrated spectrum was used as a reference laser intensity to normalize the TPE spectra. The spectra were recorded for excitation wavelengths between 650 nm to 860 nm with data acquired every 10 nm. The setup was calibrated against a coumarin 540 (Exciton) solution, which exhibits similar TPE and OPE spectra. The laser power dependence for TPE is given in the supplementary material. The TPE absorption cross section for **FR0**-SB in methanol is estimated to be 6 GM at 800 nm and 24 GM at its maximum at ~770 nm.

## III. COMPUTATIONAL DETAILS

In order to provide further insights into the enhancement of ESPT between **FR0**-SB and alcohol solvent observed in the case of TPE *vs* OPE, we augmented our experimental effort by quantum chemistry computations examining the electronic structure of the solvated **FR0**-SB system in the ground ($S_0$) and first excited singlet $S_1$ states involved in the ESPT process.[21] We focused on analyzing the role of solvation effects on the $S_0$–$S_1$ vertical and adiabatic transition energies and vertical transition dipole moments, along with the electronic dipoles characterizing the individual $S_0$ and $S_1$ states of **FR0**-SB, which are key quantities in comparing the one- and two-photon $S_0 \rightarrow S_1$ absorption cross sections. In doing so, we relied on the coupled-cluster (CC) theory,[38] which provides an accurate and size-extensive description of molecular systems, and its extension to excited states using the equation-of-motion (EOM) CC formalism,[39] focusing on the EOMCC approach with singles and doubles (EOMCCSD)[39] and the $\delta$-CR-EOMCC(2,3) triples correction[40] to EOMCCSD, which is a rigorously size-intensive modification to the CR-EOMCC(2,3)[41,42] methodology capable of determining excitation energies to within ~0.1–0.2 eV.[43] In modeling the solvated **FR0**-SB chromophore, we considered the complex of **FR0**-SB hydrogen-bonded to a cluster of three alcohol solvent molecules, designated as [**FR0**-SB⋯HOR], which, according to our previous investigation of the steric effects on the ESPT process involving **FR0**-SB and *n*- and *i*-propanol, is the minimum number of explicit solvent molecules required for the proton transfer to occur.[37] Following Ref. 37, we used the "branched" arrangement of the three alcohol solvent molecules treated in our modeling explicitly, with one of them hydrogen-bonded to **FR0**-SB and the other two solvating it, since such an arrangement leads to the lowest energy barriers for the ESPT reactions involving **FR0**-SB (see Ref. 37 for further details). The remaining, bulk, solvation effects were described using the continuum solvation model based on the solute electron density (SMD) approach.[44] The alcohol solvents considered in our computations were methanol, ethanol, *n*-propanol, and *i*-propanol.

For each of the alcohol solvents considered in our calculations, the geometry optimization of the [**FR0**-SB···HOR] complex in its $S_0$ state, used in the subsequent CC/EOMCC calculations, was performed using density functional theory (DFT)[45] employing the Kohn-Sham formulation of DFT.[46] To obtain the corresponding minimum-energy structures of the [**FR0**-SB···HOR] species in the $S_1$ state, we used the time-dependent (TD)[47] extension of DFT to excited electronic states. In carrying out these geometry optimizations, we used the CAM-B3LYP functional[48] which, as elaborated on in our earlier studies,[21,37] provides vertical excitation energies of **FR0**-SB that are closer to those resulting from the EOMCC calculations using the $\delta$-CR-EOMCC(2,3) triples correction to EOMCCSD than the excitation energies obtained with other tested functionals. All geometry optimizations of the [**FR0**-SB···HOR] complex employed the 6-31+G* basis set[49–51] and accounted for the bulk solvation effects using the aforementioned SMD model.

To provide accurate information about the transition energies and transition dipole moments characterizing the absorption ($S_0 \rightarrow S_1$) and emission ($S_1 \rightarrow S_0$) processes involving the solvated **FR0**-SB species and the corresponding dipoles in the $S_0$ and $S_1$ states, which are all needed to model the one- and two-photon cross sections for each of the alcohol solvents considered in our calculations, we performed the following series of single-point CC and EOMCC computations at the aforementioned CAM-B3LYP/6-31+G*/SMD optimized geometries. First, we determined the $S_0$–$S_1$ electronic transition energies

$$\omega_{10}^{(\text{EOMCC})} = E_{S_1}^{(\text{EOMCC})} - E_{S_0}^{(\text{CC})} \tag{1}$$

corresponding to the [**FR0**-SB···HOR] complex in the absence of the SMD continuum solvation, where the total electronic energies of the $S_0$ and $S_1$ states entering Eq. (1) were computed as

$$E_{S_0}^{(\text{CC})} = E_{S_0}^{(\text{CCSD/6-31+G*})} + [E_{S_0}^{(\text{CR-CC}(2,3)/6\text{-}31G)} - E_{S_0}^{(\text{CCSD/6-31G})}] \tag{2}$$

for the ground state and

$$E_{S_1}^{(\text{EOMCC})} = E_{S_1}^{(\text{EOMCCSD/6-31+G*})} + [E_{S_1}^{(\delta\text{-CR-EOMCC}(2,3)/6\text{-}31G)} - E_{S_1}^{(\text{EOMCCSD/6-31G})}] \tag{3}$$

for the first excited singlet state. The first term on the right-hand side of Eq. (2) denotes the total electronic energy of the $S_0$ state computed at the CCSD[52] level utilizing the largest basis set considered in this study, namely, 6-31+G*. The term in the square brackets on the right-hand side of Eq. (2) corrects the CCSD/6-31+G* energy for the many-electron correlation effects due to triply excited clusters obtained in the CR-CC(2,3)[41,53,54] calculations employing the smaller and more affordable 6-31G basis.[49] Similarly, the first term on the right-hand side of Eq. (3) designates the EOMCCSD/6-31+G* energy of the $S_1$ state and the expression in the square brackets represents the triples correction to EOMCCSD obtained in the $\delta$-CR-EOMCC(2,3)/6-31G calculations. Ideally, one would like to use basis sets larger than 6-31+G* and, in particular, incorporate polarization and diffuse functions on hydrogen atoms, but such calculations at the CC and EOMCC levels used in this work turned out to be prohibitively expensive. Nevertheless, we tested the significance of the polarization[50] and diffuse[51] functions on hydrogen atoms by performing the CAM-B3LYP/6-31++G**/SMD calculations for the [**FR0**-SB···HOR] complexes which show that neither the excitation energies nor the dipole and transition dipole moment values change by more than 1 % compared to the CAM-B3LYP/6-31+G*/SMD results.

Before describing the remaining elements of our computational protocol, it is important to emphasize that the composite approach defined by Eqs. (1)–(3) is more general than the analogous expressions shown in Ref. 21 where we focused on the vertical excitation processes only. Equations (1)–(3) encompass both the vertical and adiabatic transition energies. Indeed, if $E_{S_0}^{(CC)}$ and $E_{S_1}^{(EOMCC)}$ are calculated at the minimum on the $S_0$ potential energy surface, $\omega_{10}^{(EOMCC)}$ given by Eqs. (1)–(3) becomes the vertical excitation energy $\omega_{10}^{(EOMCC)}(\text{abs.})$ characterizing the $S_0 \to S_1$ absorption defined by Eq. (1) of Ref. 21. If $E_{S_0}^{(CC)}$ and $E_{S_1}^{(EOMCC)}$ are determined at the minimum characterizing the [**FR0**-SB···HOR] complex in the $S_1$ state, we obtain the vertical transition energy $\omega_{10}^{(EOMCC)}(\text{em.})$ corresponding to the $S_1 \to S_0$ emission. The $\omega_{10}^{(EOMCC)}$ energy defined by Eq. (1) becomes the adiabatic transition energy, abbreviated as $\omega_{10}^{(EOMCC)}(\text{ad.})$, when $E_{S_0}^{(CC)}$ and $E_{S_1}^{(EOMCC)}$ are computed at their respective minima. As far as the transition dipole moments characterizing the vertical absorption and emission processes involving the solvated **FR0**-SB species are concerned, they were calculated from the one-electron transition density matrices obtained at the EOMCCSD level of theory employing the 6-31+G* basis set. Similarly, we used the CCSD/6-31+G* and EOMCCSD/6-31+G* one-electron reduced density matrices to determine the dipole moments of the $S_0$ and $S_1$ states at each of the two potential minima.

Given the large computational costs associated with the EOMCCSD and $\delta$-CR-EOMCC(2,3) calculations for the [**FR0**-SB···HOR] system, which consists of three alcohol molecules bound to the **FR0**-SB chromophore and which requires correlating as many as 216 electrons and 758 molecular orbitals in the case of the *n*- or *i*-propanol solvents when the 6-31+G* basis set is employed, we replaced the three explicit alcohol molecules with the corresponding effective fragment potentials (EFPs).[55] We were able to do this because, based on our CAM-B3LYP/6-31+G*/SMD calculations for the [**FR0**-SB···HOR] complexes, the $S_0$–$S_1$ electronic transition does not involve charge transfer between the photobase and its solvent environment. Indeed, the $S_0$–$S_1$ transition in the bare[21] and solvated **FR0**-SB species has a predominantly π–π* character with the π and π* orbitals localized on the **FR0**-SB chromophore, *i.e.*, the alcohol solvent molecules are mere spectators to this excitation process (see the supplementary material for further details). The use of EFPs to represent the cluster of three alcohol molecules bonded to **FR0**-SB in our CC/EOMCC computations allowed us to reduce the system size to that of the bare **FR0**-SB species embedded in the external potential providing a highly accurate description of the intermolecular interactions between **FR0**-SB and solvent molecules in the [**FR0**-SB···HOR] complex, including electrostatic, polarization, dispersion, and exchange repulsion effects.[55]

Once the electronic transition energies and the corresponding one-electron properties of the [**FR0**-SB···HOR] complex were determined, we proceeded to the second stage of our modeling protocol, which was the incorporation of the remaining bulk solvation effects that turned out to be non-negligible as well. As in the case of the aforementioned geometry optimizations, the bulk solvation effects were calculated with the help of the implicit solvation SMD approach. Due to limitations of the computer codes available to us, we could not perform the CC/EOMCC computations in conjunction with the SMD model, so we estimated the SMD effects using the *a posteriori* corrections $\delta_X^{(SMD)}$ to the various CC/EOMCC properties $X$ of the [**FR0**-SB···HOR] complex, such as transition energies and dipole moments, using DFT and TD-DFT. These corrections were constructed in the following way. First, for each of the four alcohol solvents considered in our calculations, we performed single-point DFT/TD-DFT calculations for the [**FR0**-SB···HOR] complex at the previously optimized $S_0$ and $S_1$ geometries accounting for the bulk solvation effects using SMD. As in the case of the geometry optimizations, we used the CAM-B3LYP functional and the 6-31+G* basis set and, in analogy to the CC/EOMCC computations, replaced the cluster of three explicit alcohol solvent molecules bound to **FR0**-SB by the corresponding EFPs. We then repeated the analogous calculations without SMD. This allowed us to determine the desired $\delta_X^{(SMD)}$ corrections using the formula

$$\delta_X^{(SMD)} = X^{(CAM-B3LYP/6-31+G*/SMD)} - X^{(CAM-B3LYP/6-31+G*)}, \qquad (4)$$

where the first and second terms on the right-hand side of Eq. (4) designate property $X$ obtained in the CAM-B3LYP/6-31+G* calculations with and without SMD, respectively. The final SMD-corrected EOMCC electronic transition energies were computed as

$$\omega_{10} = \omega_{10}^{(EOMCC)} + \delta_{\omega_{10}}^{(SMD)}, \qquad (5)$$

where $\omega_{10}^{(EOMCC)}$ is the transition energy for the [**FR0**-SB⋯HOR] complex defined by Eqs. (1)–(3), whereas the SMD-corrected one-electron properties were determined using the formula

$$X = X^{[(EOM)CCSD/6-31+G*]} + \delta_X^{(SMD)}, \qquad (6)$$

with $X^{[(EOM)CCSD/6-31+G*]}$ denoting the value of property $X$ calculated at the (EOM)CCSD/6-31+G* level. If the property of interest was a vector, such as dipole or transition dipole moment, we used Eq. (6) for each of the Cartesian components of the vector.

Finally, to gauge the effects of solvation on the various calculated properties, including transition energies and dipole and transition dipole moments, we also performed single-point CC/EOMCC calculations for the bare super photobase, *i.e.*, **FR0**-SB without the presence of explicit solvent molecules or equivalent EFPs and SMD implicit solvation, at the gas-phase geometry of the $S_1$ state optimized using CAM-B3LYP/6-31+G*. In the case of the $S_0$ minimum-energy structure, we relied on our previous gas-phase CC/EOMCC results reported in Ref. 21.

All of the electronic structure calculations reported in this work, including the CAM-B3LYP geometry optimizations with and without the SMD continuum solvation, the CC/EOMCC single-point calculations without implicit SMD solvation, and the CAM-B3LYP single-point calculations with and without SMD, needed to estimate the SMD corrections to CC/EOMCC properties, were performed using the GAMESS package[56,57] (we used the 2019 R2 version of GAMESS). In the case of the $S_0 \rightarrow S_1$ absorption process, whenever the SMD implicit solvation model was utilized, we incorporated the nonequilibrium solvation effects associated with the solvent relaxation delay, as implemented in GAMESS.[58] The relevant CCSD, CR-CC(2,3), EOMCCSD, and $\delta$-CR-EOMCC(2,3) computations using the restricted Hartree-Fock (RHF) determinant as a reference and the corresponding left-eigenstate CCSD and EOMCCSD calculations, which were needed to determine the triples corrections of CR-CC(2,3) and $\delta$-CR-EOMCC(2,3) and the one-electron properties of interest, including the dipole and transition dipole moments, were carried out using the CC/EOMCC routines developed by the Piecuch group,[41,42,53,54,59–61] which form part of the GAMESS code as well. In all of our CC/EOMCC calculations, the core orbitals associated with the 1s shells of C and N atoms of **FR0**-SB were kept frozen. The EFPs that were used to replace the cluster of three explicit alcohol solvent molecules bound to **FR0**-SB in the CC/EOMCC single-point calculations and the CAM-B3LYP computations aimed at determining the SMD solvation effects were generated using the RHF approach and the 6-31+G* basis set. Thanks to the use of EFPs, our frozen-core CC/EOMCC calculations for the [**FR0**-SB···HOR] complex correlated only 138 electrons of the **FR0**-SB system. In all of the calculations employing the 6-31+G* basis set, we used spherical components of d orbitals. The Cartesian coordinates of all structures used in the quantum chemistry calculations reported in this work, along with the resulting $S_0$ and $S_1$ dipole moment vectors, can be found in the supplementary material. Visual representations of the structures and their $S_0$ and $S_1$ dipoles are provided in the supplementary material as well.

## IV. RESULTS

Steady-state fluorescence spectra following isoenergetic OPE (400 nm) and TPE (2 × 800 nm) for a number of alcohols are presented in Fig. 2. These spectra have been divided by the frequency cubed according to the transition dipole moment representation, which makes fluorescence intensity proportional to population according to the Einstein coefficient of spontaneous emission.[62] Upon excitation, **FR0**-SB reaches the first excited singlet state (denoted **FR0**-SB*), which emits at ~21,000 cm$^{-1}$. Following proton transfer, the **FR0**-HSB$^{+*}$ excited protonated state is reached, which emits at ~15,000 cm$^{-1}$. The probability of proton transfer is observed to decrease as the alkane chain of the linear alcohols increases. This observation was found to correlate with the relative –OH concentration.[21] Interestingly, for *i*-propanol, significantly less proton transfer takes place, an aspect related to steric hindrance in the formation of an appropriate solvent configuration for proton transfer that has been addressed elsewhere.[37]

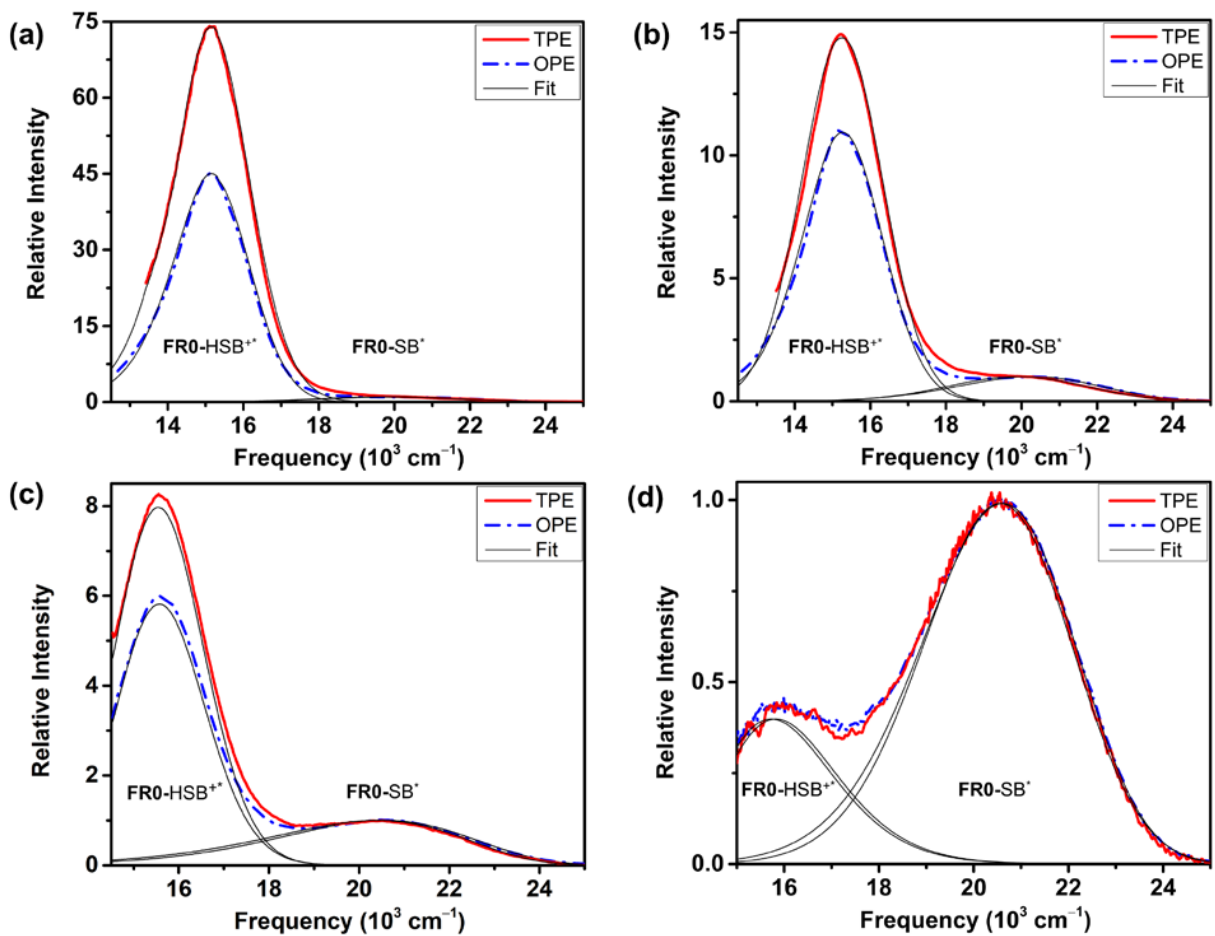

**FIG. 2**. OPE and TPE steady-state fluorescence spectra obtained for **FR0**-SB in (a) methanol, (b) ethanol, (c) *n*-propanol, and (d) *i*-propanol. In each of the panels, OPE (blue line) is compared with TPE (red line). The fluorescence spectra are normalized to the non-protonated emission intensity. The ratio between the areas for **FR0**-HSB$^{+*}$ (~15,000 cm$^{-1}$) and **FR0**-SB* (~21,000 cm$^{-1}$) emission following OPE and TPE is determined by fits to log-normal functions (thin black lines).

Of particular interest in this work is how the probability for proton transfer, *i.e.*, the reactivity of the Schiff base, is affected by the excitation process. For this purpose, we quantify the extent of proton transfer as the [**FR0**-HSB$^{+*}$]/[**FR0**-SB*] ratio for linear and nonlinear excitation by fitting the fluorescence areas to log-normal functions as shown in Fig. 2, and then correcting the results for differences in the fluorescence quantum yield of the protonated and non-protonated excited-state species in the different solvents. The results, summarized in Table I, are presented in Fig. 3 for both OPE (blue) and TPE (red); note that the vertical axis on the right is a logarithmic scale so differences between the two modes of excitation seem less prominent than they actually are. While both OPE and TPE proton-transfer rates change proportionally for the different solvents, we consistently observe greater proton transfer following TPE compared to OPE. The ratio between the two modes of observed excitation, *i.e.*, the ratio of ratios, is shown as gray bars in Fig. 3. We find that for methanol TPE leads to 62 ± 20 % greater reactivity than OPE. Greater reactivity following TPE *vs* OPE is also observed for ethanol (42 ± 13 %), *n*-propanol (36 ± 4 %), and *n*-hexanol (24 ± 3 %), although the percent enhancement decreases with alcohol aliphatic chain length. The large error bars, especially for methanol, result from the difficulty in measuring the very small **FR0**-SB* signal, which amounts to one part in 58. In the case of *i*-propanol, no excess reactivity is found within the uncertainty of the measurements.

**TABLE I**. Quantitative assessment of the extent of protonation following OPE and TPE from steady-state fluorescence measurements for **FR0**-SB. The ratio of the extent of protonation expressed as TPE/OPE shows the enhancement in ESPT resulting from TPE experiments compared to their OPE counterparts.

| Solvent[a] | OPE ratio | TPE ratio | TPE/OPE |
|---|---|---|---|
| MeOH | 36 ± 3 | 58 ± 5 | 1.62 ± 0.20 |
| EtOH | 7.6 ± 0.5 | 10.9 ± 0.7 | 1.42 ± 0.13 |
| *n*-PrOH | 3.85 ± 0.06 | 5.2 ± 0.1 | 1.36 ± 0.04 |
| *n*-HxOH | 1.72 ± 0.02 | 2.13 ± 0.04 | 1.24 ± 0.03 |
| *i*-PrOH | 0.37 ± 0.01 | 0.35 ± 0.02 | 0.94 ± 0.06 |

[a] Abbreviations: MeOH = methanol, EtOH = ethanol, *n*-PrOH = *n*-propanol, *n*-HxOH = *n*-hexanol, *i*-PrOH = *i*-propanol.

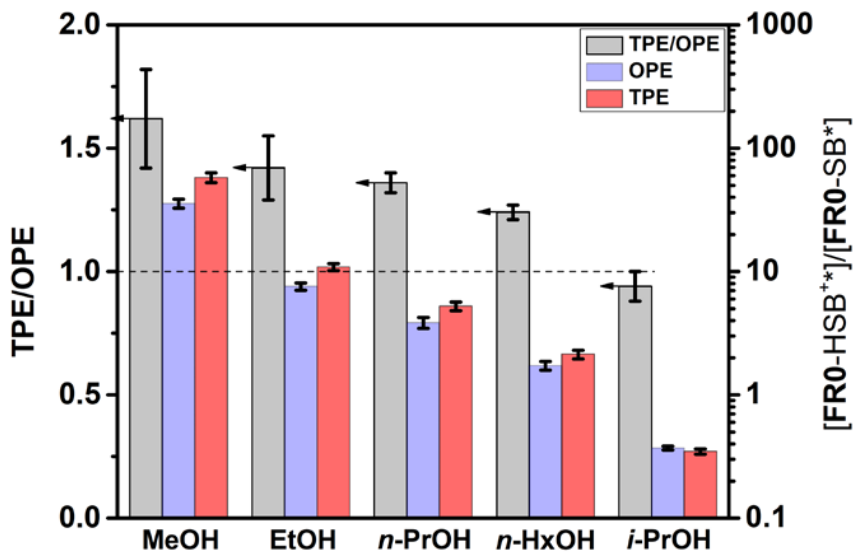

**FIG. 3**. Extent of proton transfer determined following OPE and TPE for **FR0**-SB in a number of alcohols. The TPE over OPE ratio is given by the gray bars (left y-axis). In all the solvents but *i*-propanol, enhanced proton transfer is observed following TPE. The ratios between the fluorescence bands [**FR0**-HSB$^{+*}$]/[**FR0**-SB*], corrected for fluorescence quantum yield, are plotted as bars OPE (blue) and TPE (red); each ratio is indicated on a logarithmic scale (right y-axis). Abbreviations: MeOH = methanol, EtOH = ethanol, *n*-PrOH = *n*-propanol, *n*-HxOH = *n*-hexanol, *i*-PrOH = *i*-propanol.

The enhanced reactivity following isoenergetic TPE is unexpected. Therefore, we explore the possible involvement of an additional dark state that lies within the absorption band associated with $S_1$ or a higher excited singlet state $S_n$ that is reached via three-photon excitation ($3 \times 800$ nm). These possible contributions are addressed as follows. Excitation–emission matrix (EEM) spectra were obtained for **FR0**-SB in methanol and ethanol and are shown in Fig. 4. The absorption spectrum is shown as a bold black line. From the spectra in Fig. 4, we observe no evidence for the existence of an additional state within the 325–450 nm $S_1$ region. However, we do see evidence for absorption to $S_n$ with $n > 1$ in the 225–290 nm region which is also associated with **FR0**-HSB$^{+*}$ and **FR0**-SB* emission following $S_n$ to $S_1$ internal conversion. We measured the excitation intensity dependence of the TPE integrated fluorescence for the different solvents (see Fig. S1 of the supplementary material). The exponent associated with the laser intensity dependence indicates the number of photons associated with the excitation process. The exponent measured was ~1.9, indicating that three-photon excitation, if it occurs, contributes minimally. Thus, to summarize, we exclude the participation of a dark state near $S_1$ based on the EEM spectra. Furthermore, given the near quadratic laser power dependence combined with the observation that the probability for proton transfer following 266 nm excitation is similar to that following 400 nm excitation, contributions to the observed ESPT enhancement from three-photon excitation processes seem unlikely. While we cannot rule out the involvement of excited-state absorption in which two-photon $S_0 \rightarrow S_1$ transition is followed by a one-photon $S_1 \rightarrow S_n$ excitation in the observed reactivity enhancement, we note that direct excitation at 266 nm does not lead to enhanced reactivity and, when normalizing for proton transfer emission, we find that maximum proton transfer for OPE is observed at 400 nm excitation (see Fig. S3 of the supplementary material).

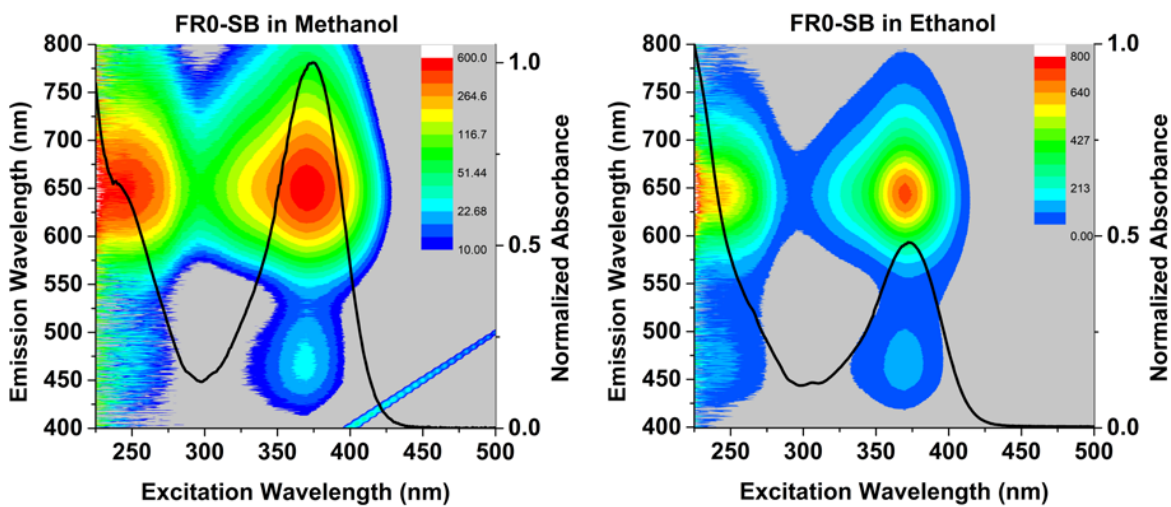

**FIG. 4**. EEM spectra showing the dependence of protonation as a function of excitation wavelength for methanol (left) and ethanol (right). The absorption spectrum for both molecules is shown as a bold black line. Emission from **FR0**-SB* is observed at ~450 nm and emission from **FR0**-HSB$^{+*}$ is observed at ~650 nm.

The striking solvent dependence observed for the enhanced two-photon reactivity implies that the underlying process depends on the dynamics of proton transfer and solvation. To explore this dependence, we can first rely on our previous time-resolved TCSPC results for the solvents studied here.[21] In Fig. 5, we plot the TPE/OPE enhancement as a function of the measured **FR0**-SB* lifetime. We observe an inverse correlation, shorter **FR0**-SB* lifetimes correlate with greater enhancement. When the **FR0**-SB* lifetime is longer, the observed enhancement decreases. We interpret this finding as follows. Although **FR0**-SB has a high p$K_a$*, the ability of **FR0**-SB* to abstract a proton depends on the solvent configuration. We know from our quantum chemistry calculations reported in Ref. 37 that the minimum of three solvent molecules, with one of them directly hydrogen-bonded to **FR0**-SB, are needed to enable the ESPT process. Achieving such a configuration is easiest for small molecules, such as methanol, and much less probable for secondary alcohols, *e.g.*, *i*-propanol and cyclopentanol.[37] Thus, we postulate that TPE prepares the molecule at a point on the excited-state reaction coordinate that enhances reactivity, but such propensity is lost within a few hundred picoseconds or less.

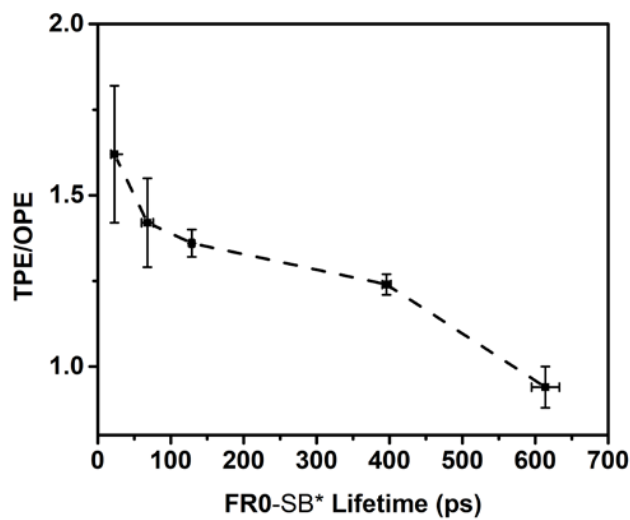

**FIG. 5**. Isoenergetic two-photon enhanced ESPT as a function of excited-state lifetime prior to proton transfer for (in order of greater to lower enhancement) methanol, ethanol, *n*-propanol, *n*-hexanol, and *i*-propanol. The dashed line is included as a guide to the eye.

From the previous observations, it appears that TPE leads to a more reactive species than OPE. We performed fluorescence lifetime measurements following isoenergetic OPE and TPE detecting at both the **FR0**-SB* and **FR0**-HSB$^{+*}$ wavelength regions. Results from these measurements are summarized in Table II. In the case of aprotic solvents, such as acetonitrile, the average excited-state lifetime $\bar{\tau}_{SB}$ is ~2 ns. The much shorter values for $\bar{\tau}_{SB}$ in alcohols are associated with the formation of the [**FR0**-SB*⋯HOR] complex where the proton is already shared by **FR0**-SB* and ROH, which precedes the separation and solvation of the protonated **FR0**-HSB$^{+*}$ and the deprotonated solvent RO$^-$ species characterized by the rise time $\tau_X$. The latter process has a timescale that is much less dependent on the method of excitation. We find that the first step in protonation, namely, the loss of population in **FR0**-SB* and the rise of the **FR0**-HSB$^{+*}$ emission are approximately two times faster for TPE than for OPE. This observation is consistent with the enhanced reactivity and with the conclusion that TPE leads to a more reactive species. Measurements carried out in acetonitrile in the **FR0**-SB* and **FR0**-HSB$^{+*}$ wavelength regions showed no OPE *vs* TPE difference, indicating that the enhancement depends on the hydrogen-bonding capabilities of protic solvents.

**TABLE II**. Fluorescence lifetime measurements following one- and two-photon isoenergetic excitation of **FR0**-SB in methanol, ethanol, and acetonitrile. The initial state, **FR0**-SB*, decays with a fast $\tau_{SB1}$ and a slow $\tau_{SB2}$ biexponential lifetimes. The numbers in parentheses indicate the amplitude of the fast decay component ($a_1$). The protonated state **FR0**-HSB$^{+*}$ shows a fast rise time $\tau_X$ and a slow decay time $\tau_{HSB}$.[37] In acetonitrile, no proton transfer takes place, thus one observes only a single exponential decay of the **FR0**-SB* state that is identical for OPE and TPE within the measurement errors. All numbers are given in picoseconds.

| Solvent[a] | OPE | | | | | TPE | | | | |
|---|---|---|---|---|---|---|---|---|---|---|
| | $\tau_{SB1}$ | $\tau_{SB2}$ | $\bar{\tau}_{SB}$[b] | $\tau_X$ | $\tau_{HSB}$ | $\tau_{SB1}$ | $\tau_{SB2}$ | $\bar{\tau}_{SB}$[b] | $\tau_X$ | $\tau_{HSB}$ |
| MeOH | 23 ± 7 (0.99) | 1259 ± 14 | 35 ± 7 | 31 ± 1 | 1116 ± 2 | 12 ± 6 (0.99) | 811 ± 30 | 20 ± 6 | 28 ± 9 | 1106 ± 11 |
| EtOH | 59 ± 4 (0.84) | 220 ± 6 | 85 ± 4 | 103 ± 2 | 1241 ± 33 | 35 ± 3 (0.89) | 212 ± 20 | 54 ± 5 | 90 ± 4 | 1232 ± 1 |
| ACN | — | 2121 ± 10 | 2121 ± 10 | — | — | — | 2138 ± 8 | 2138 ± 8 | — | — |

[a] Abbreviations: MeOH = methanol, EtOH = ethanol, ACN = acetonitrile.
[b] For protic solvents (MeOH and EtOH), $\bar{\tau}_{SB} = a_1 \tau_{SB1} + a_2 \tau_{SB2}$, where $a_2 = 1 - a_1$. In the case of ACN, for which there is no ESPT, $\bar{\tau}_{SB} = \tau_{SB2}$.

Having ruled out the involvement of an additional excited state, or excitation to a higher $S_n$ excited state with $n > 1$, we now turn to the possibility of reaching a more reactive species via TPE. We begin by comparing the expressions for the absorption cross sections associated with OPE and TPE arising from the first- and second-order time-dependent perturbation theory, respectively (see, *e.g.*, Ref. 63). The $0 \rightarrow f$ OPE absorption cross section, with 0 and $f$ denoting the initial and final electronic states, respectively, is[63]

$$\sigma^{(1)}_{f0}(\omega) = A |\mu_{f0}|^2 g_{M1}(\omega), \tag{7}$$

where $\omega$ is the frequency of the exciting photon (in our case, the frequency of a 400 nm laser), $A$ is a constant, $\mu_{f0}$ denotes the magnitude of the transition dipole moment between the ground and excited electronic states, and $g_{M1}(\omega)$ is the OPE distribution function or linewidth associated with the molecular system of interest. In presenting Eq. (7), we have assumed an isotropic averaging over the directions of the transition dipole moment vector $\boldsymbol{\mu}_{f0}$. To arrive at an expression for the absorption cross section associated with the isoenergetic one-color TPE, where the laser frequency is half of its OPE counterpart, we take advantage of the fact that no resonance at 800 nm is observed in our experiments, in agreement with our electronic structure calculations. Under these conditions, the absorption cross section for TPE becomes[64]

$$\sigma^{(2)}_{f0}(\omega/2) = B \left| \sum_v \frac{\mu_{fv} \mu_{v0}}{\omega_{v0} - \omega/2 + i\Gamma_v(\omega/2)} \right|^2 g_{M2}(\omega), \tag{8}$$

where $B$ is a constant, $\omega_{v0}$ is the frequency needed to reach the intermediate state $v$ from the ground state 0, $i\Gamma_v(\omega/2)$ is a damping factor that is inversely proportional to the lifetime of a given intermediate state $v$, and $g_{M2}(\omega)$ is the TPE line shape function. In analogy to the OPE absorption cross section, we have performed an isotropic averaging over the directions of the transition dipole moment vectors $\boldsymbol{\mu}_{fv}$ and $\boldsymbol{\mu}_{v0}$.

Equation (8) is useful, but in this work we are interested in relating the TPE absorption cross section with the change in the dipole moment upon $0 \to f$ photoexcitation. One can derive such a relationship if we perform the following mathematical manipulations.[65] First, we separate the $v = 0$ and $v = f$ terms from the sum over states in Eq. (8). Next, we take advantage of the fact that in our case 0 and $f$ correspond to the electronically bound $S_0$ and $S_1$ states of **FR0**-SB, respectively. This allows us to eliminate the $i\Gamma_v(\omega/2)$ term in the $v = 0$ and $v = f$ denominators in Eq. (8). In the final step, we replace $\omega_{f0}$ in the $v = f$ denominator by $\omega$ and combine the $v = 0$ and $v = f$ contributions to obtain[65]

$$\sigma^{(2)}_{f0}(\omega/2) = B \left| \sum_{v \neq 0, f} \frac{\mu_{fv}\mu_{v0}}{\omega_{v0} - \omega/2 + i\Gamma_v(\omega/2)} + \frac{\mu_{f0}\Delta\mu_{f0}}{\omega/2} \right|^2 g_{M2}(\omega). \tag{9}$$

It is customary to refer to the first term in Eq. (9) as the 'virtual' pathway and to the second one, which relies on the transition dipole moment $\mu_{f0}$ and the difference between the permanent ground- and excited-state dipoles $\Delta\mu_{f0} \equiv \mu_{ff} - \mu_{00}$, as the 'dipole' pathway.[66] Equation (9) shows that for centrosymmetric molecules, for which $\Delta\mu_{f0}$ vanishes identically, the virtual pathway is the only contributing term to the TPE absorption cross section. However, **FR0**-SB is not centrosymmetric and, thus, it is interesting to examine the extent to which each pathway contributes to the $S_0 \to S_1$ one-color TPE considered here. For the first term in Eq. (9) to be large, the following three conditions would have to be satisfied: (1) the $\omega_{v0}$ frequency characterizing the $0 \to v$ transition would have to be close to the frequency $\omega/2$ of each of the two photons associated with TPE, (2) the $i\Gamma_v(\omega/2)$ damping factor would have to be very small, i.e., the intermediate state $v$ would have to be sufficiently long-lived, and (3) the $0 \to v$ and $v \to f$ transitions would have to be allowed, giving rise to larger $\mu_{v0}$ and $\mu_{fv}$ transition dipole moments. In the case of the TPE experiments performed in this work, it is unlikely that conditions (1), (2), and (3) can be simultaneously satisfied. Indeed, since there are no dipole-allowed electronic states between $S_0$ and $S_1$, the intermediate state $v$ satisfying condition (1) would have to be a rovibrational resonance supported by the ground-state electronic potential. It is unlikely that such resonances are long-lived and characterized by large $0 \rightarrow v$ and $v \rightarrow f$ transition dipole moments. It is possible that the intermediate states $v$ characterized by larger $\mu_{v0}$ and $\mu_{fv}$ values exist, but those would have to be electronic states higher in energy than $S_1$, which cannot satisfy the resonant condition (1). Furthermore, as demonstrated in Ref. 21, the low-lying electronically excited states above $S_1$ are characterized by small or even negligible transition dipole moments from the ground state. In other words, while the virtual pathway contributes to the TPE cross section, the probability that it dominates it seems low, especially when we realize that there are reasons for the dipole pathway to play a substantial role in the case of the molecular systems considered in this work. Indeed, as shown in our earlier studies,[21,37] and as further elaborated on below, the $S_0 \rightarrow S_1$ excitations in the isolated and solvated **FR0**-SB are characterized by large transition dipole moments and significant changes in the permanent dipoles. This suggests that the second term in Eq. (9) plays a major role, which is consistent with the well-established fact that the dipole pathway becomes critical when TPE involves charge transfer associated with substantial change in the permanent dipole upon photoexcitation.[67–73] Although the $S_0 \rightarrow S_1$ transition in **FR0**-SB is accompanied by a migration of a small amount of charge,[21,37] this migration happens over a very large distance, giving rise to more than a three-fold increase in dipole moment and a substantial enhancement of the second term in Eq. (9). Given the above analysis, from this point on we focus on the dipole pathway and assume that we can approximate the TPE absorption cross section by the second term in Eq. (9), i.e.,[70]

$$\sigma_{f0}^{(2)}(\omega/2) \approx B' \left|\mu_{f0}\right|^2 \left|\Delta\mu_{f0}\right|^2 g_{M2}(\omega), \tag{10}$$

where $B' = 4B/\omega^2$.

As illustrated in Eqs. (7) and (10), the absorption cross sections for both one- and two-photon excitation processes depend on the square of the absolute value of the transition dipole moment $\mu_{f0}$ characterizing the $0 \rightarrow f$ vertical electronic excitation, which in our case is the transition dipole $\mu_{10}$ corresponding to the $S_0 \rightarrow S_1$ photoabsorption for the **FR0**-SB system in various solvents. However, in the case of TPE, the absorption cross section also depends on the difference between the electronic dipole moments of the $f$ and 0 states, $\Delta\mu_{f0}$, which in our case is the difference $\Delta\mu_{10} \equiv \mu_1 - \mu_0$ between the dipole moment $\mu_1$ characterizing the first excited singlet S$_1$ state of **FR0**-SB and its S$_0$ counterpart $\mu_0$. Consequently, $\Delta\mu_{10}$ and its dependence on the solvent environment hold the key to understanding the enhancement of the ESPT reactions between the **FR0**-SB photobase and alcohol solvents observed in the case of TPE. To provide insights into the effect of solvation on $\Delta\mu_{10}$ and other properties characterizing the S$_0$ and S$_1$ states of the solvated **FR0**-SB chromophore and transitions between them, we performed electronic structure calculations using the CC/EOMCC-based composite approach described in the Computational Details section. As mentioned in that section, the alcohol solvents considered in our computations were methanol, ethanol, *n*-propanol, and *i*-propanol.

In Table III, we report the vertical transition energies $\omega_{10}$(abs.) and transition dipole moments $\mu_{10}$ characterizing the S$_0$ → S$_1$ photoabsorption process, along with the dipoles corresponding to the S$_0$ and S$_1$ states, $\mu_0$ and $\mu_1$, respectively, and their ratios resulting from our calculations for **FR0**-SB in the gas phase and in the aforementioned four solvents determined at the minima on the respective S$_0$ potential energy surfaces. The analogous information for the S$_1$ → S$_0$ emission and the dipole moment values of the S$_0$ and S$_1$ states determined at the S$_1$ minima characterizing the isolated and solvated **FR0**-SB is presented in Table IV. We begin our discussion of computational results by comparing the vertical absorption and emission energies characterizing the solvated **FR0**-SB species obtained with the CC/EOMCC-based protocol adopted in this work against their experimental counterparts. The vertical excitation energies for the [**FR0**-SB⋯HOR] complexes calculated at the respective S$_0$ minima, shown in Table III, are essentially identical to the locations of the peak maxima in the corresponding experimental photoabsorption spectra reported in Ref. 37, which are 3.32, 3.33, 3.32, and 3.34 eV for methanol, ethanol, *n*-propanol, and *i*-propanol, respectively. The same accuracies are also seen in the case of the vertical emission energies calculated at the S$_1$ minima of the [**FR0**-SB⋯HOR] species reported in Table IV, which can hardly be distinguished from the maxima in the experimental emission peaks for **FR0**-SB in methanol, ethanol, *n*-propanol, and *i*-propanol of 2.57, 2.61, 2.62, and 2.65 eV, respectively.[37] These observations corroborate the accuracy of the computational protocol used in this study to model the interactions of the **FR0**-SB photobase with the various alcohol solvents. The observed good agreement between the theoretical vertical transition energies reported in Tables III and IV and the corresponding experimental data can largely be attributed to the use of high-level *ab initio*

CC/EOMCC approaches in describing the [**FR0**-SB···HOR] complexes. This becomes apparent when one considers the errors relative to experiment characterizing the vertical transition energies obtained in the single-point CAM-B3LYP/6-31+G*/SMD computations, which are about 0.2–0.3 eV (9–11 %).

**TABLE III**. The vertical transition energies $\omega_{10}$(abs.) (in eV) and transition dipole moments $\mu_{10}$ (in Debye) corresponding to the $S_0 \rightarrow S_1$ absorption, along with the $\mu_0$ and $\mu_1$ dipoles characterizing the $S_0$ and $S_1$ states (in Debye) and their ratios for **FR0**-SB in the gas phase and in selected alcohol solvents calculated at the respective $S_0$ minima following the CC/EOMCC-based protocol described in Computational Details.

| Solvent[a] | $\omega_{10}$(abs.) | $\mu_{10}$ | $\mu_0$ | $\mu_1$ | $\mu_1/\mu_0$ |
|---|---|---|---|---|---|
| None (gas phase)[b] | 3.70 | 6.9 | 2.6 | 8.6 | 3.3 |
| MeOH | 3.30 | 9.6 | 4.4 | 16.4 | 3.8 |
| EtOH | 3.32 | 9.5 | 4.3 | 15.9 | 3.7 |
| $n$-PrOH | 3.32 | 9.5 | 4.2 | 15.9 | 3.7 |
| $i$-PrOH | 3.33 | 9.4 | 4.2 | 15.7 | 3.7 |

[a] Abbreviations: MeOH = methanol, EtOH = ethanol, $n$-PrOH = $n$-propanol, $i$-PrOH = $i$-propanol.
[b] Taken from our previous gas-phase CC/EOMCC calculations reported in Ref. 21.

**TABLE IV**. The vertical transition energies $\omega_{10}$(em.) (in eV) and transition dipole moments $\mu_{10}$ (in Debye) corresponding to the $S_1 \rightarrow S_0$ emission, along with the $\mu_0$ and $\mu_1$ dipoles characterizing the $S_0$ and $S_1$ states (in Debye) and their ratios for **FR0**-SB in the gas phase and in selected alcohol solvents calculated at the respective $S_1$ minima following the CC/EOMCC-based protocol described in Computational Details.

| Solvent[a] | $\omega_{10}$(em.) | $\mu_{10}$ | $\mu_0$ | $\mu_1$ | $\mu_1/\mu_0$ |
|---|---|---|---|---|---|
| None (gas phase) | 3.26 | 8.9 | 3.4 | 10.9 | 3.2 |
| MeOH | 2.68 | 11.8 | 6.6 | 20.0 | 3.0 |
| EtOH | 2.69 | 11.8 | 6.5 | 19.7 | 3.0 |
| $n$-PrOH | 2.70 | 11.8 | 6.5 | 19.6 | 3.0 |
| $i$-PrOH | 2.72 | 11.7 | 6.4 | 19.1 | 3.0 |

[a] Abbreviations: MeOH = methanol, EtOH = ethanol, $n$-PrOH = $n$-propanol, $i$-PrOH = $i$-propanol.

Having established the accuracy of our quantum chemistry protocol, we proceed to the discussion of our computational findings regarding the dipole moments of the $S_0$ and $S_1$ states and the transition dipoles between them, which are the key quantities for the one- and two-photon absorption cross sections given by Eqs. (7) and (10), respectively. In the absence of direct experimental information, our computations provide insights into the effects of solvation on these quantities. To begin with, as reported in our earlier work for the bare **FR0**-SB species,[21] and as shown in Table III, there is a large, by a factor of more than 3, increase in the electronic dipole moment following $S_0 \rightarrow S_1$ photoabsorption, giving rise to the superbase character of **FR0**-SB*. Upon solvation, both $S_0$ and $S_1$ dipole moments of the **FR0**-SB chromophore are significantly enhanced, becoming approximately twice as large as their gas-phase counterparts. This can be attributed to the polarization of the electron cloud of the **FR0**-SB photobase by the alcohol molecules surrounding it. Furthermore, the fact that the electronic dipole moment characterizing the $S_1$ state is much larger than its $S_0$ counterpart translates into a stronger stabilization of the $S_1$ state relative to $S_0$, leading to lower $S_0 \rightarrow S_1$ vertical excitation energies in the case of **FR0**-SB in alcohol solvents when compared to the bare **FR0**-SB system. The transition dipole moment characterizing the $S_0 \rightarrow S_1$ photoabsorption process is amplified by solvation as well (by about 40 %), which results in larger OPE and TPE absorption cross sections for the solvated **FR0**-SB species relative to their gas-phase values. Similar trends are observed when we examine the dipoles and transition dipoles shown in Table IV. It is also interesting to note that the dipole moments characterizing the $S_0$ and $S_1$ states and the corresponding transition dipoles increase upon geometrical relaxation from the $S_0$ to $S_1$ minima, with a concomitant red shift in the vertical transition energies. This bathochromic shift is more pronounced in the case of the solvated **FR0**-SB species as a consequence of $\mu_1$ being much larger than $\mu_0$, implying a stronger stabilization of the $S_1$ state due to the polar solvent environment compared to the $S_0$ state.

As already alluded to above, the transition dipole moments characterizing the $S_0$–$S_1$ absorption and emission processes and the $S_0$ and $S_1$ dipoles at the respective potential minima could not be determined from our experiments. However, by analyzing the solvatochromic shift of the absorption and fluorescence bands in sixteen different solvents as a function of solvent dielectric constant and index of refraction, we could estimate the magnitude of the transition dipole moment $\mu_{10}$ and the change in the dipole moment, $\Delta\mu_{10}$, associated with the $S_0 \rightarrow S_1$ adiabatic excitation.[74,75] Based on our analysis, we found $\Delta\mu_{10}$ of **FR0**-SB in the alcohol solvents considered in our experiments to be ~15 Debye, a magnitude usually associated with substantial charge transfer, and $\mu_{10}$ to be about 10 Debye. The procedure outlining how the experimental values of $\mu_{10}$ and $\Delta\mu_{10}$ were derived is given in the supplementary material. Having access to the dipole moments characterizing the $S_0$ and $S_1$ states at their respective minimum-energy structures and the vertical transition dipole moments associated with the $S_0$–$S_1$ transitions resulting from our quantum chemistry computations (see Tables III and IV) allowed us to assess the quality of our experimentally derived values of $\mu_{10}$ and $\Delta\mu_{10}$. As shown in Tables III and IV, the vertical transition dipole moments $\mu_{10}$ characterizing the **FR0**-SB chromophore in the alcohol solvents included in our calculations range from 9.4 to 11.8 Debye, in very good agreement with the experimentally derived value of about 10 Debye. According to the data collected in Table V, the calculated and experimentally derived changes in the dipole moment associated with the $S_0 \to S_1$ adiabatic transition, which are about 15 Debye in both cases, are virtually identical. Given that both theory and experiment point to the large values of $\mu_{10}$ and $\Delta\mu_{10}$ as a result of solvation and that the dipole pathway defined by the second term in Eq. (9) is anticipated to be the dominant TPE pathway, as discussed above, we can conclude that using Eq. (10) in approximating the TPE absorption cross section of **FR0**-SB in alcohol solvents is justified.

**TABLE V**. A comparison of the calculated $S_0$–$S_1$ adiabatic transition energies without [$\omega_{10}$(ad.)] and with [$\omega_{10}$(0-0)] zero-point energy (ZPE) vibrational corrections (in eV), along with the differences and ratios of the $\mu_0$ and $\mu_1$ dipoles characterizing the $S_0$ and $S_1$ states at the respective minima (in Debye) for **FR0**-SB in the gas phase and in selected alcohol solvents obtained following the CC/EOMCC-based protocol described in Computational Details with the corresponding experimentally derived data.

| | Theory | | | | Experiment | | | |
|---|---|---|---|---|---|---|---|---|
| Solvent[a] | $\omega_{10}$(ad.) | $\omega_{10}$(0-0)[b] | $\Delta\mu_{10}$[c] | $\mu_1/\mu_0$[c] | Solvent[a] | $\omega_{10}$(0-0) | $\Delta\mu_{10}$[d] | $\mu_1/\mu_0$[d] |
| None (gas phase) | 3.42 | 3.33 | 8.3 | 4.2 | $c$-Hexane | 3.4 | — | — |
| MeOH | 2.88 | 2.80 | 15.6 | 4.6 | MeOH | 2.9 | 15.2 ± 0.2 | 4.4 ± 0.1 |
| EtOH | 2.89 | 2.80 | 15.4 | 4.6 | EtOH | 3.0 | 15.3 ± 0.3 | 4.6 ± 0.1 |
| $n$-PrOH | 2.89 | 2.81 | 15.3 | 4.6 | $n$-PrOH | 3.0 | 15.3 ± 0.3 | 4.6 ± 0.1 |
| $i$-PrOH | 2.88 | 2.80 | 14.9 | 4.5 | $i$-PrOH | 3.0 | 15.5 ± 0.5 | 4.7 ± 0.1 |

[a] Abbreviations: MeOH = methanol, EtOH = ethanol, $n$-PrOH = $n$-propanol, $i$-PrOH = $i$-propanol, $c$-Hexane = cyclohexane.
[b] Calculated as $\omega_{10}$(ad.) + $\Delta$ZPE, where $\Delta$ZPE is the difference between the zero-point vibrational energies characterizing the $S_1$ and $S_0$ electronic states of the bare **FR0**-SB molecule in the gas phase computed at the CAM-B3LYP/6-31+G* level of theory. Our calculations with and without solvent indicate that the effect of solvation on $\Delta$ZPE is negligible (less than 0.01 eV).
[c] Calculated using the $\mu_0$ values reported in Table III and the $\mu_1$ values reported in Table IV.
[d] Calculated using the theoretical values of $\mu_0$ reported in Table III and the procedure based on the analysis of the experimental solvatochromic shifts described in the supplementary material.

By forming the ratio of Eqs. (7) and (10), we can obtain a new expression that summarizes the difference between OPE and TPE, in which the change in permanent dipole moment acts as an amplification factor,

$$\frac{\sigma_{f0}^{(2)}(\omega/2)}{\sigma_{f0}^{(1)}(\omega)} = \frac{B'}{A} \frac{|\Delta\mu_{f0}|^2 |\mu_{f0}|^2 g_{M2}(\omega)}{|\mu_{f0}|^2 g_{M1}(\omega)} = C \frac{|\Delta\mu_{f0}|^2 g_{M2}(\omega)}{g_{M1}(\omega)}, \qquad (11)$$

where $C = B'/A$. The difference between OPE and TPE typically arises from differences in the expressions for the line shapes, which in the case of large organic molecules in solution, are the Franck-Condon distribution convolved with the extensive homogeneous and inhomogeneous broadening. We have acquired these spectra for **FR0**-SB in cyclohexane, acetonitrile, and methanol, as shown in Fig. 6. From these spectra, we can obtain the ratio of the spectral line shapes for OPE and TPE.

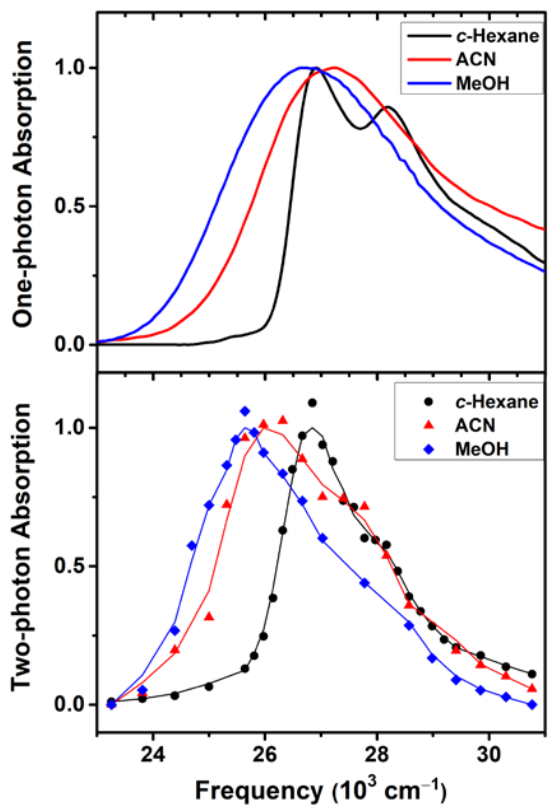

**FIG. 6**. OPE and TPE spectra for **FR0**-SB in cyclohexane (*c*-Hexane), acetonitrile (ACN), and methanol (MeOH). The solid line shown for the two-photon spectra is the result of 3-point smoothing and is included as a guide to the eye.

For cyclohexane, a non-polar solvent for which the change in permanent dipole moment is smallest, the excitation maximum for OPE and TPE coincide, and the ratio between the line shapes can be fit to a line with a negative slope from the peak to higher energies, indicating a steeper decline of the spectrum as a function of excitation energy for TPE. The steeper decline is consistently observed for acetonitrile and for methanol. However, for acetonitrile and methanol we also see that the two-photon absorption peaks appear at significantly lower energies than their OPE counterparts. The largest shift of the TPE absorption maximum compared to OPE, of ~1,140 cm$^{-1}$, is observed for methanol. Differences between OPE and TPE have been observed experimentally, in particular as blue shifts in the TPE of the green fluorescent protein,[76] and have been explained as non-Condon contributions to TPE.[77] However, the very large red shift observed in this work appears to be unprecedented.

Having strong evidence that TPE must be reaching parts of the $S_1$ potential surface that enhance reactivity compared to the isoenergetic OPE and that the initial state wave packet following TPE is different than that following OPE, we turn our attention to the relationship between the ESPT enhancement, the inhomogeneous broadening, and the significant red shift observed in our TPE experiments. It is well-established that polar solvents, in particular those capable of forming hydrogen bonds, are responsible for significant inhomogeneous broadening in absorption spectra.[78,79] The inhomogeneous broadening can further be amplified when the solute has hydrogen-bond acceptor and/or donor functional groups and undergoes large permanent dipole changes upon photoexcitation.[78] From Eq. (10), we learn that TPE is greatly enhanced by the $\left|\Delta\mu_{f0}\right|^2$ term, favoring chromophore molecules whose local solvation environment gives rise to larger $\Delta\mu_{f0}$ values compared to OPE. The substantial increase in the dipole moment upon photoexcitation should result in the different configurations in which protic solvents can be arranged to solvate **FR0**-SB, giving rise to inhomogeneous broadening. At the same time, larger $\Delta\mu_{f0}$ results in the additional stabilization of the [**FR0**-SB···HOR] complex in the $S_1$ state relative to the ground state, manifesting itself in the observed red shift in the absorption maximum. While this hypothesis needs additional thorough investigations using, for example, two-dimensional spectroscopy and molecular dynamics simulations of the observed excited-state reactivity, it is conceivable that the dynamical restructuring of the solvent around the chromophore molecules following TPE, promoting larger $\Delta\mu_{f0}$ values compared to OPE, results in stronger charge transfer between the amine and imine nitrogens of **FR0**-SB, the additional accumulation of negative charge on the imine nitrogen, and, subsequently, the enhancement of the ESPT process, which is what we attempted to schematically illustrate in Fig. 1(b). This is in contrast to some intramolecular ESPT reactions, such as in diethylaminohydroxyflavone, where solvation is inversely correlated with proton transfer.[80] The differences observed in **FR0**-SB following OPE or TPE are caused by the large change in permanent dipoles influencing the probability for TPE of some molecules.

Molecules with a solvent configuration that is more likely to result in proton transfer would then be favored to undergo TPE because of their larger $\left|\Delta\mu_{f0}\right|^2$ values.

We find support for the role of inhomogeneous broadening in protic solvents in the experimental data. We note that there is essentially no shift between the maxima for OPE and TPE for cyclohexane, a non-polar molecule. There is a ~1,000 cm$^{-1}$ shift observed for acetonitrile, and a ~1,140 cm$^{-1}$ shift for methanol. These shifts are not predicted simply by the non-Condon vibrational makeup of the excited-state wave packet. The existence of differently solvated species, in particular those exhibiting greater dipole moment changes, can be determined experimentally by comparing the absorption spectra for **FR0**-SB in solvents that exhibit TPE enhancement (methanol, ethanol, *n*-propanol) with that of **FR0**-SB in *i*-propanol, which does not. The comparison is shown in Fig. 7, where we plot the absorption spectra as well as the difference between the absorption spectrum in each solvent and that in *i*-propanol. We observe a large difference in the absorption spectra, especially at 25,000 cm$^{-1}$, where the experiment was carried out. The largest difference is found for methanol, which exhibits the largest TPE enhancement.

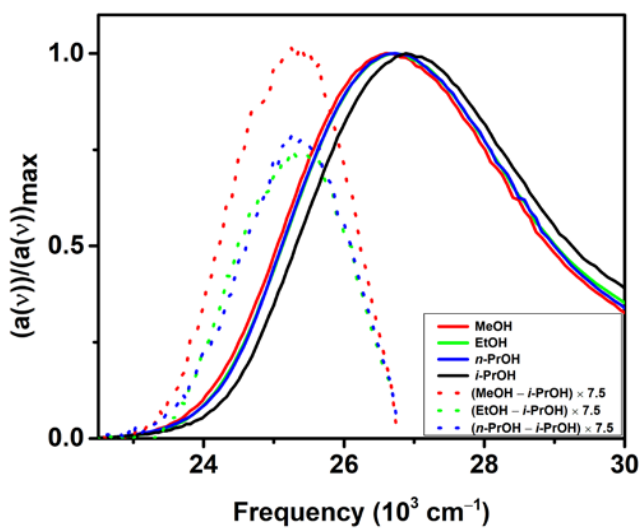

**FIG. 7**. Absorption spectra of **FR0**-SB in methanol (MeOH), ethanol (EtOH), *n*-propanol (*n*-PrOH), and *i*-propanol (*i*-PrOH) (solid lines) and difference between these spectra and the absorption in *i*-propanol (dotted lines). We find a significant difference between the spectra in the 25,000 cm$^{-1}$ energy region where the experiment was performed.

From Fig. 7, we can see significant inhomogeneous broadening toward lower energies. This supports our hypothesis that there is a clear bias toward TPE in the case of molecules primed for proton transfer, as seen in the reduction of the protonation time in methanol and ethanol by a factor of two compared to the other alcohol solvents. The effect becomes attenuated for *n*-hexanol, given that the long alkyl chain reduces the chances for two or more hydroxyl moieties to be next to the imine group where proton transfer takes place. As shown in our computations and OPE experiments reported in Ref. 37, the solvent configuration leading to proton transfer in *i*-propanol is much less likely, resulting in very low efficiency of the ESPT process, and, thus, TPE can no longer enhance it, confirming our hypothesis.

## V. CONCLUSIONS

In this paper we have presented the observation of up to 62 % enhancement in the reactivity of the super photobase **FR0**-SB following isoenergetic two-photon excitation. We have reported evidence that excitation is to the first excited singlet state, $S_1$, and that no other excited state contributes to proton transfer via one- or multi-photon excitation. We have found that the magnitude of the enhancement in reactivity correlates with the protonation rate; thus, it is faster for methanol, but slows down with longer chain alcohols and is not measurable for *i*-propanol.

We have reached the conclusion that the enhanced solvent-to-solute ESPT reactivity is the result of TPE creating a more reactive species than OPE. This is supported by analysis of the molecular properties that contribute to the one- and two-photon transition probabilities, namely, the transition dipole moment $\mu_{10}$ coupling the $S_0$ and $S_1$ states of **FR0**-SB and the $\Delta\mu_{10}$ difference between the dipole moments of these two states. According to our mathematical manipulations, the ratio of the TPE and OPE absorption cross sections depends on $\Delta\mu_{10}$ but not on $\mu_{10}$. The $\Delta\mu_{10}$ values resulting from the high-level CC/EOMCC-based computations performed in this study were practically identical to those estimated from our experiments. We learned that the very large change in permanent dipoles between the ground and excited states of **FR0**-SB amplifies differences in the spectroscopic line shape, leading to non-Condon contributions and different vibrational makeups of the initial excited-state wave packet.

Finally, we have considered inhomogeneous broadening as providing an additional aspect leading to different excited-state species accessed via TPE. The permanent dipole change is highly dependent on the arrangement of polar solvent molecules, especially for protic solvents capable of hydrogen bonding. These added contributions are confirmed by the red shift observed in the TPE spectra for acetonitrile and methanol, but not for cyclohexane, and by the greater inhomogeneous broadening toward the low energy region in the OPE spectra of solutions exhibiting TPE enhancement. The findings reported here help explain the observation by Tokumura and Itoh on the intramolecular proton transfer in 7-hydroxyquinoline in methanol solution via two-photon excitation.[81] In that study, two-photon excitation in the 210–250 nm equivalent wavelength region leads to exclusive emission from the proton transfer state. Later measurements, based on step-wise solvation supersonic-jet spectroscopy,[82] identified that bridging methanol structures with three methanol molecules help facilitate the proton transfer.[31] These two separate observations support our conclusion that two-photon excitation favors structures primed for excited-state proton transfer.

In conclusion, we have reported unprecedented 62 % enhancement in photochemical reactivity upon isoenergetic two-photon excitation compared to one-photon excitation. The long-term goal of our research is to develop tools for precision chemistry and our findings indicate that two-photon excitation not only provides spatial and temporal control, but it can also bring about enhanced reactivity. Future work will include quantitative determination of the two-photon absorption cross section for **FR0**-SB in different solvents and of newly developed compounds.

## SUPPLEMENTARY MATERIAL

See the supplementary material for the laser power dependence of two-photon excited fluorescence, ratios between **FR0**-HSB$^{+*}$ and **FR0**-SB* emissions, fitting of the steady-state fluorescence results, time-correlated single-photon counting fitting, experimental determination of permanent dipole moments, optimized geometries and electronic dipole moments resulting from quantum chemistry computations for the bare and solvated **FR0**-SB system in the $S_0$ and $S_1$ electronic states along with their visual representations, and dominant orbitals defining the $S_0$–$S_1$ transitions in the [**FR0**-SB⋯HOR] complexes.

## ACKNOWLEDGMENTS

The collaboration between synthesis, theory, and experiments for the understanding and development of super photoreagents for precision chemistry is funded by a seed grant from DARPA and AMRDEC (W31P4Q-20-1-0001). Partial support comes from NIH (Grant Nos. 2R01EY016077-08A1 and 5R01EY025383-02 R01 to GJB, and R01GM101353 to BB), NSF (Grant No. CHE1836498 to MD) and the U.S. DOE (Grant No. DE-FG02-01ER15228 to PP). This work was supported in part through computational resources and services provided by the Institute

## DATA AVAILABILITY

The data that support the findings of this study are available within the article and its supplementary material. Further data are available from the corresponding authors upon request.

[82] L.W. Peng, M. Dantus, A.H. Zewail, K. Kemnitz, J.M. Hicks, and K.B. Eisenthal, J. Phys. Chem. **91**, 6162 (1987).

Supplementary Material:

# Isoenergetic Two-Photon Excitation Enhances Solvent-to-Solute Excited-State Proton Transfer

Jurick Lahiri[1], Mehdi Moemeni[1], Jessica Kline[1], Ilias Magoulas[1], Stephen H. Yuwono[1], Maryann Laboe[2], Jun Shen[1], Babak Borhan[1,*], Piotr Piecuch[1,3,*], James E. Jackson[1,*], G. J. Blanchard[1,*], and Marcos Dantus[1,3,*#]

[1] Department of Chemistry, Michigan State University, East Lansing, MI 48824, USA
[2] Department of Chemical Engineering and Material Science, Michigan State University, East Lansing, MI 48824, USA
[3] Department of Physics and Astronomy, Michigan State University, East Lansing, MI 48824, USA

* Corresponding authors: BB email: babak@chemistry.msu.edu, tel.: +1-517-353-0501; PP email: piecuch@chemistry.msu.edu, tel.: +1-517-353-1151; JEJ email: jackson@chemistry.msu.edu, tel.: +1-517-353-0504; GJB email: blanchard@chemistry.msu.edu, tel.: +1-517-353-1105; MD email: dantus@chemistry.msu.edu, tel.: +1-517-353-1191. [#] Lead contact.

This document contains additional information about power dependence measurements, ratios between **FR0**-HSB$^{+}$* and **FR0**-SB* emissions, fitting of the steady-state fluorescence results, time-correlated single-photon counting fitting, experimental determination of permanent dipole moments, optimized geometries and electronic dipole moments resulting from quantum chemistry computations described in the main text, and dominant orbitals defining the $S_0$–$S_1$ transitions in the [**FR0**-SB⋯HOR] complexes. This additional information is accompanied by the following supplementary figures and tables:

**Fig. S1**. Laser intensity dependence of the fluorescence following two-photon excitation for **FR0**-SB in methanol and acetonitrile.

**Fig. S2**. Power dependence of the ratio between **FR0**-HSB$^{+}$* and **FR0**-SB* emissions, [**FR0**-HSB$^{+}$*]/[**FR0**-SB*], obtained for two-photon excitation at 800 nm in methanol.

**Fig. S3**. Ratios between **FR0**-HSB$^{+}$* and **FR0**-SB* emissions, [**FR0**-HSB$^{+}$*]/[**FR0**-SB*], in methanol and ethanol as functions of excitation wavelength.

**Fig. S4**. Fluorescence of **FR0**-SB in methanol as a function of wavelength and frequency following two-photon excitation with different peak intensity laser pulses.

**Fig. S5**. Lognormal fitting of the steady-state fluorescence results in methanol, ethanol, *n*-propanol, and *i*-propanol, including residuals.

**Fig. S6**. Fitting of the fluorescence decay responses of **FR0**-SB* in methanol and ethanol detected at 460 nm following one-photon and two-photon excitations, including residuals.

**Fig. S7**. The CAM-B3LYP/6-31+G* optimized structures of **FR0**-SB in its $S_0$ and $S_1$ states.

**Fig. S8**. The CAM-B3LYP/6-31+G*/SMD optimized structures of **FR0**-SB in its $S_0$ and $S_1$ states hydrogen-bonded to a cluster of three methanol solvent molecules.

**Fig. S9**. The CAM-B3LYP/6-31+G*/SMD optimized structures of **FR0**-SB in its $S_0$ and $S_1$ states hydrogen-bonded to a cluster of three ethanol solvent molecules.

**Fig. S10**. The CAM-B3LYP/6-31+G*/SMD optimized structures of **FR0**-SB in its $S_0$ and $S_1$ states hydrogen-bonded to a cluster of three *n*-propanol solvent molecules.

**Fig. S11**. The CAM-B3LYP/6-31+G*/SMD optimized structures of **FR0**-SB in its $S_0$ and $S_1$ states hydrogen-bonded to a cluster of three *i*-propanol solvent molecules.

**Fig. S12**. The HOMO and LUMO that characterize the $S_0 \to S_1$ photoexcitation in the complex of **FR0**-SB hydrogen-bonded to a cluster of three methanol solvent molecules.

**Table S1**. Results obtained from lognormal fitting of the steady-state fluorescence data.

**Table S2**. Calculated semiempirical dipole moments characterizing the $S_0$ and $S_1$ states, $\mu_0$ and $\mu_1$, respectively, along with their differences and ratios for **FR0**-SB in the selected solvents.

# Power Dependence Measurements, Ratios between FR0-HSB$^{+}$* and FR0-SB* Emissions, and Fitting of the Steady-State Fluorescence Results

Power dependence measurements were performed in methanol and acetonitrile solutions with the Ti:Sapphire oscillator that was used for most of the measurements. The laser was centered at 800 nm and the pulses were 16 fs in duration. Analysis of the power dependence resulted in a slope slightly less than 2 (*cf.* Fig. S1), which is consistent for compounds that undergo large changes in their permanent dipole moment upon excitation.

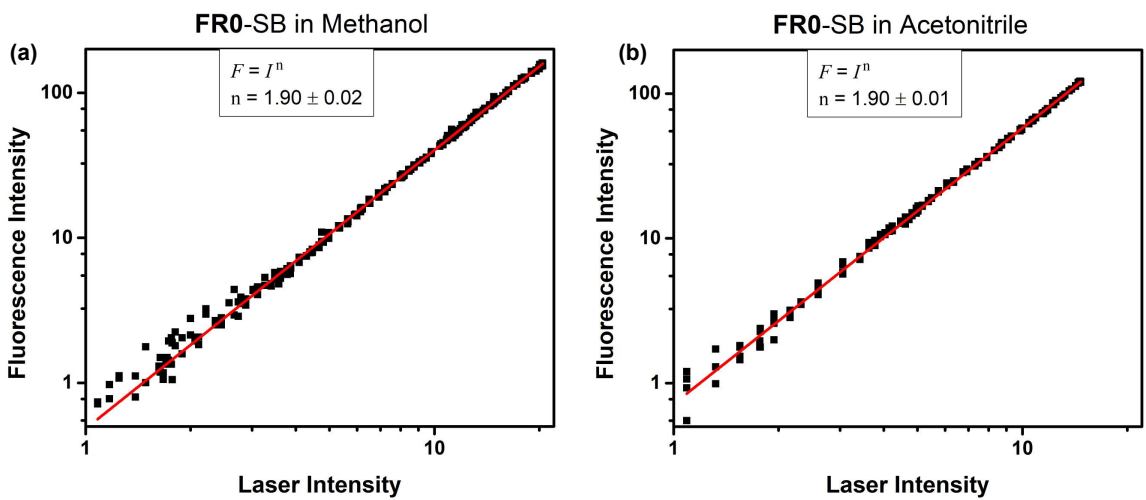

**Fig. S1**. Laser intensity dependence of the fluorescence following two-photon excitation for **FR0**-SB in (a) methanol and (b) acetonitrile, measured for laser pulses from a Ti:Sapphire oscillator. The fitting function (red line) and the experimental points (black squares) are plotted on a log-log scale. The exponent obtained from the fit, of ~2, indicates two-photon excitation and shows no indication of three-photon excitation.

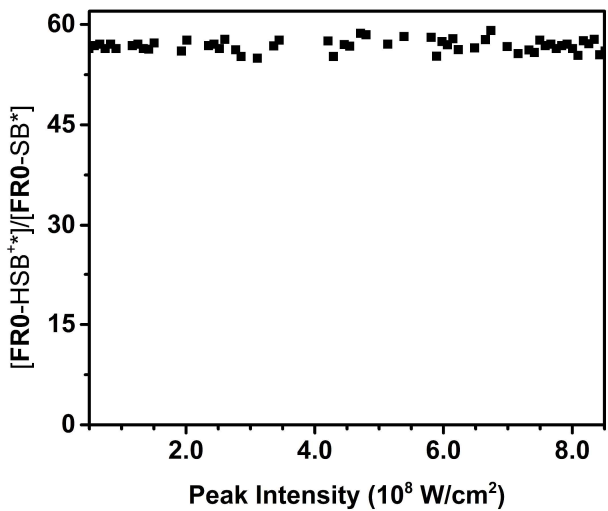

**Fig. S2**. Power dependence of the ratio between **FR0**-HSB$^{+}$* and **FR0**-SB* emissions, [**FR0**-HSB$^{+}$*]/[**FR0**-SB*], obtained for two-photon excitation at 800 nm in methanol. We find that the probability for excited-state proton transfer is independent of laser intensity. This finding excludes the contribution from three-photon or higher photonicity processes to the observed enhanced proton transfer when replacing one-photon excitation by two-photon excitation.

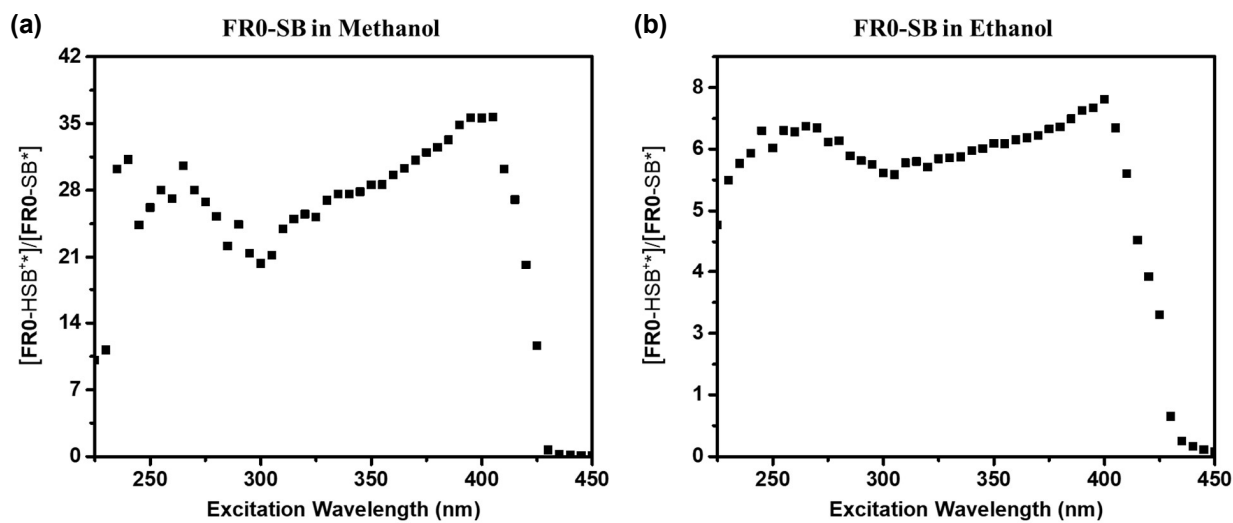

**Fig. S3**. Ratios between **FR0**-HSB$^+$* and **FR0**-SB* emissions, [**FR0**-HSB$^+$*]/[**FR0**-SB*], in (a) methanol and (b) ethanol as functions of excitation wavelength. We note that in both methanol and ethanol solutions the maximum protonation occurs at 400 nm. This indicates that three-photon excitation near 266 nm, even if it occurred, would not yield higher rate of protonation.

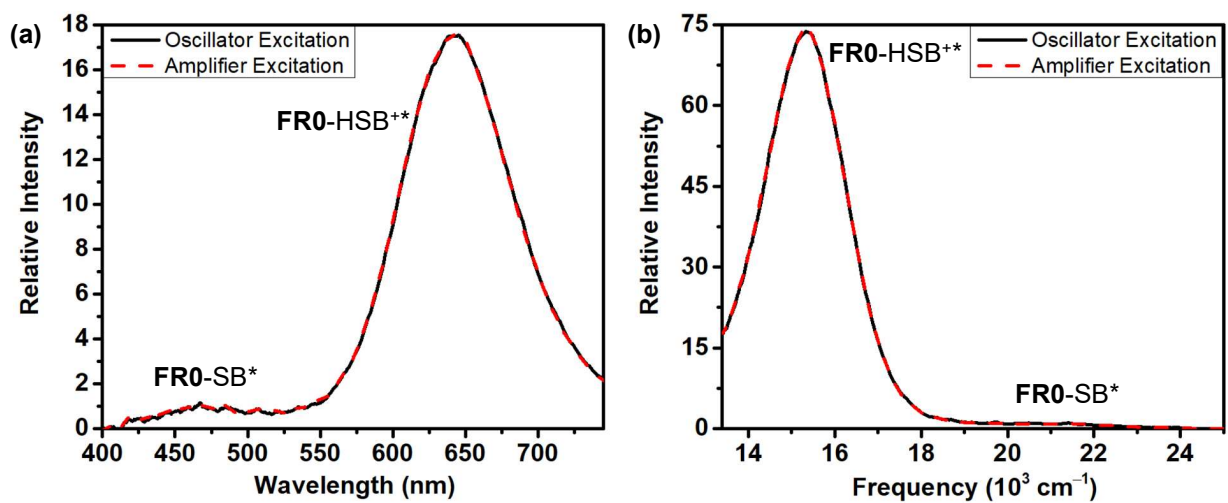

**Fig. S4**. Fluorescence of **FR0**-SB in methanol as a function of (a) wavelength and (b) frequency following two-photon excitation with different peak intensity laser pulses. The solid black curve corresponds to fluorescence arising from lower peak-power excitation of ~$10^8$ W/cm$^2$ from a Ti:Sapphire oscillator while the red dashed curve represents fluorescence following higher peak-power excitation of ~$10^{12}$ W/cm$^2$ from a Ti:Sapphire amplifier. All spectra are normalized to the emission of **FR0**-SB*. In panel (b), the intensity has been divided by the frequency cubed in order to obtain amplitudes that are proportional to population (see the main text). These data show that the probability for excited-state proton transfer (corresponding to emission of **FR0**-HSB$^+$* near 650 nm) is not affected by the peak power of the laser.

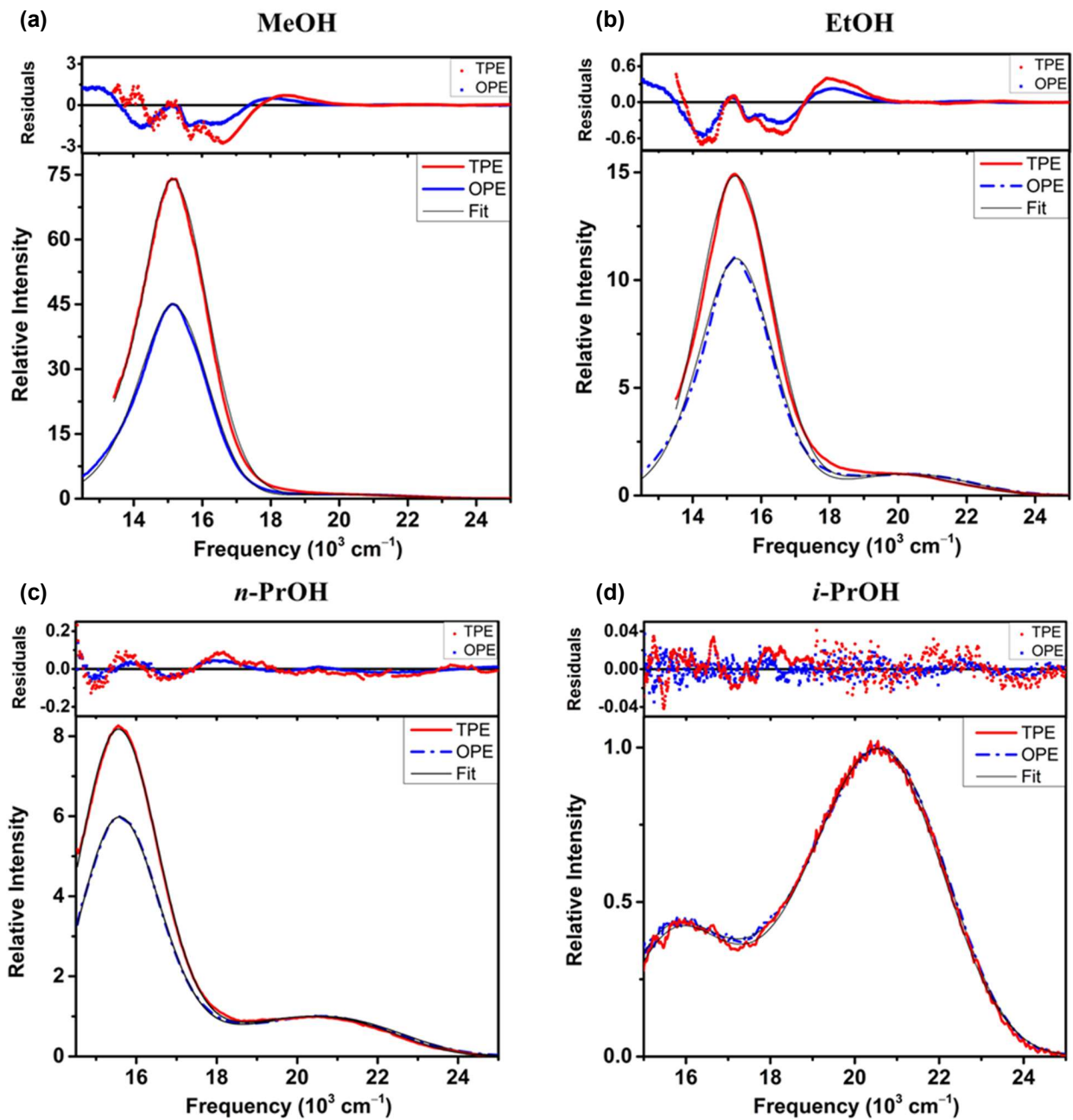

**Fig. S5**. Lognormal fitting of the steady-state fluorescence results in (a) methanol, (b) ethanol, (c) *n*-propanol, and (d) *i*-propanol, including residuals. These spectra have been divided by the frequency cubed according to the transition dipole moment representation, which makes fluorescence intensity proportional to population according to the Einstein coefficient of spontaneous emission. The residuals show no systematic bias to the ratio between **FR0**-HSB$^+$* and **FR0**-SB* emissions.

**Table S1**. Results obtained from lognormal fitting of the steady-state fluorescence data according to the formula below[a] for the spectra shown in Fig. S5. The fitting has been carried out with the data normalized to the emission maxima corresponding to different alcohol solvents. The extent of protonation following OPE and TPE resulting from these fits are reported in Table I of the main text.

| Solvent | FR0-SB* | | | | FR0-HSB$^{+}$* | | | |
|---|---|---|---|---|---|---|---|---|
| | $h_1$ | $\gamma_1$ | $v_{01}$ | $\Delta_1$ | $h_2$ | $\gamma_2$ | $v_{02}$ | $\Delta_2$ |
| | | | | OPE | | | | |
| MeOH | 2.2E−02 ± 5.0E−04 | −1.0E−03 ± 6.5E−05 | 20032 ± 85 | 4508 ± 134 | 1.00 ± 0.02 | −1.0E−01 ± 1.3E−03 | 15150 ± 10 | 2420 ± 23 |
| EtOH | 9.0E−02 ± 1.9E−04 | 1.1E−08 ± 3.2E−10 | 20050 ± 58 | 5017 ± 170 | 0.99 ± 0.02 | 3.2E−09 ± 9.6E−11 | 15220 ± 50 | 2465 ± 30 |
| n-PrOH | 1.7E−01 ± 5.5E−04 | −2.5E−01 ± 5.4E−03 | 20468 ± 15 | 5013 ± 18 | 0.97 ± 0.01 | 6.1E−02 ± 9.5E−04 | 15575 ± 1 | 2303 ± 1 |
| i-PrOH | 9.8E−01 ± 2.1E−03 | −1.6E−01 ± 9.6E−03 | 20612 ± 8 | 3982 ± 31 | 0.37 ± 0.01 | 6.0E−03 ± 5.9E−05 | 15724 ± 10 | 2729 ± 46 |
| | | | | TPE | | | | |
| MeOH | 1.3E−02 ± 4.0E−04 | 1.0E−02 ± 2.1E−07 | 20010 ± 98 | 4425 ± 223 | 1.00 ± 0.02 | 1.0E−01 ± 3.3E−03 | 15150 ± 50 | 2360 ± 70 |
| EtOH | 6.7E−02 ± 1.5E−04 | 9.4E−10 ± 2.8E−11 | 20010 ± 89 | 4850 ± 100 | 0.99 ± 0.02 | 8.7E−03 ± 5.9E−05 | 15230 ± 76 | 2496 ± 30 |
| n-PrOH | 1.2E−01 ± 7.6E−04 | −3.0E−01 ± 1.4E−03 | 20478 ± 43 | 5085 ± 69 | 0.96 ± 0.01 | 4.4E−02 ± 1.4E−04 | 15557 ± 2 | 2322 ± 2 |
| i-PrOH | 9.7E−01 ± 2.9E−03 | −1.6E−01 ± 1.4E−02 | 20553 ± 9 | 3913 ± 33 | 0.37 ± 0.01 | 3.3E−02 ± 6.7E−04 | 15786 ± 16 | 2488 ± 67 |

[a] The formula used in the fitting is $L_i(v) = h_i \exp\left(-(\ln 2)\left[\frac{\ln(1+2\gamma_i(v-v_{0i})/\Delta_i)}{\gamma_i}\right]^2\right)$.

# Time-Correlated Single-Photon Counting Fitting

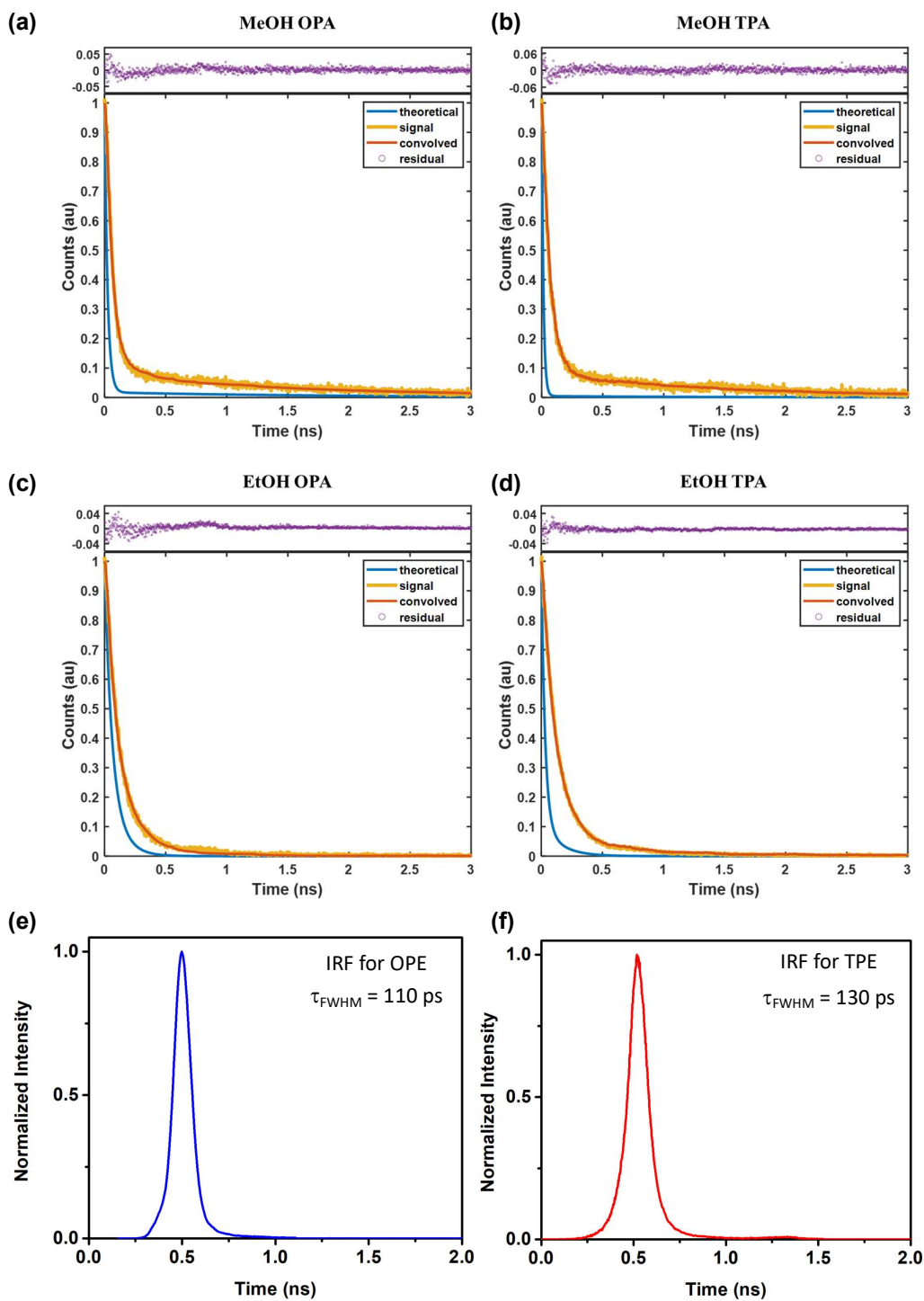

**Fig. S6**. Fitting of the fluorescence decay responses of **FR0**-SB* in methanol [panels (a) and (b)] and ethanol [panels (c) and (d)] detected at 460 nm following one-photon [panels (a) and (c)] and two-photon [panels (b) and (d)] excitations, including residuals. The fitting function used was $f(t) = a_1 \exp(-t/\tau_{SB1}) + a_2 \exp(-t/\tau_{SB2})$. For the time constants reported in the main text, the instrument response function (IRF) has been deconvoluted using a convolute-and-compare method. The results of the fitting (labeled as "convolved") correspond to the fitting function (labeled as "theoretical") convoluted with the IRF shown in panels (e) and (f).

# Experimental Calculation of Permanent Dipole Moments

A number of theories have been developed to correlate the solvatochromic shift exhibited by fluorescent polar molecules upon solvation to the change in the permanent dipole moments between the ground (*g*) and excited (*e*) electronic states, $\Delta\mu_{eg} \equiv \mu_e - \mu_g$. Typically, analysis tabulates the solvatochromic shift for a series of solvents and these trends are used to determine the ratio of the permanent dipole moments $\mu_e/\mu_g$ (see Ref. 75 of the main text). We followed this approach for 16 different solvents (methanol, ethanol, *n*-propanol, *i*-propanol, *n*-butanol, *n*-pentanol, *n*-hexanol, *n*-octanol, ethylene glycol, dimethyl sulfoxide, dimethylformamide, acetone, dichloromethane, ethyl acetate, diethyl ether, *c*-hexane) and obtained $\mu_e/\mu_g = 4.4 \pm 0.8$ for **FR0**-SB. When we only took into account aprotic solvents, we obtained $\mu_e/\mu_g = 3.7 \pm 0.6$. The uncertainty in these numbers reflects how the compound behaves in different solvents given that some are protic and some aprotic, some are polar and some are not. Our interest is to determine how **FR0**-SB behaves in the different solvents studied here, hence we need more granular information. To that end, we use a semiempirical approach that borrows some of the theoretical values reported in the main text. We start from expressions given for the solvatochromic shift (see Ref. 62 of the main text) and shown in Eq. (S1):

$$\tilde{\nu}_{abs} - \tilde{\nu}_{00} = -\frac{2}{hca^3}\mu_g(\mu_e - \mu_g)f,$$

$$\tilde{\nu}_{flr} - \tilde{\nu}_{00} = -\frac{2}{hca^3}\mu_e(\mu_e - \mu_g)f,$$

(S1)

where *a* is the Onsager cavity radius and *f* represents a solvent parameter resulting from second-order perturbation theory that takes into account the Onsager reaction field for a spherical cavity of radius *a* and depends on the permittivity $\varepsilon$ and the index of refraction of the solvent *n*. These expressions apply for solvents with large *f*, which causes a large solvatochromic shift. For solvents that have very small *f*, such as alkanes with $f \sim 0$, these expressions are not applicable. For an isotropic polarizability of the solute, *f* becomes

$$f(\varepsilon, n) = \frac{2n^2+1}{n^2+2}\left(\frac{\varepsilon-1}{\varepsilon+2} - \frac{n^2-1}{n^2+2}\right).$$

(S2)

We note that it is easy to find the absorption and fluorescence maxima entering Eq. (S1), $\tilde{\nu}_{abs}$ and $\tilde{\nu}_{flr}$, respectively, but not the 0-0 transition origin, $\tilde{\nu}_{00}$. All the constants before the dipole moments in Eq. (S1) are also hard to determine with accuracy and carry a number of assumptions such as the cavity radius being independent of solvent, the molecule being spherical, and the ground- and excited-state dipole moments being aligned. However, if we subtract the bottom part from the top part of Eq. (S1), we obtain

$$\tilde{\nu}_{abs} - \tilde{\nu}_{flr} = \alpha\mu_e(\mu_e - \mu_g)f - \alpha\mu_g(\mu_e - \mu_g)f,$$

(S3)

where we have replaced all the constants by $\alpha$. We notice that the left-hand side is the solvatochromic shift *S* from the peak of the absorption to the peak of the fluorescence spectra, which is easily determined, while on the right-hand side we find the difference between ground- and excited-state permanent dipole moments. We now obtain a much simpler expression

$$S = \alpha f(\mu_e - \mu_g)^2 = \alpha f \Delta\mu_{eg}^2.$$

(S4)

Note that the solvatochromic shift $S$, $f$, and the permanent dipole moments are solvent-dependent. Experimentally, we cannot get the permanent dipole moments with the needed accuracy. However, we can take the ground-state permanent dipole from theory. In that case, we can use the experimental data to calculate the excited-state permanent dipole, the ratio of permanent dipole moments, and the change in dipole,

$$(\mu_e - \mu_g)^2 = S/\alpha f,$$
$$(\mu_e - \mu_g) = \sqrt{S/\alpha f}, \quad \text{(S5)}$$
$$\mu_e = \sqrt{S/\alpha f} + \mu_g.$$

We can now calculate the values for the excited-state dipole moment, as well as the change in permanent dipole for each of the solvents. The only parameter missing is $\alpha$, which can be determined globally for all the solvents. For this determination, we used the top part of Eq. (S5) using the experimental solvatochromic shift values, and the $f$ parameter for each solvent. We used the theoretical values of the change in permanent dipole moments for this calculation. For each solvent we found a slightly different $\alpha$ value. We then took the average as the global solution used to calculate all the individual parameters using the bottom part of Eq. (S5), and the standard deviation was used to calculate the uncertainties. We find that one $\alpha$ value leads to very consistent results for all four alcohols, indicating that the approximate semiempirical approach is valid. A similar calculation for $c$-hexane could not be performed because this method is not applicable for non-polar solvents with $f \sim 0$.

**Table S2.** Calculated semiempirical dipole moments characterizing the $S_0$ and $S_1$ states, $\mu_0$ and $\mu_1$, respectively, (in Debye) along with their differences and ratios for **FR0**-SB in the selected solvents. The values in green shade were obtained following the CC/EOMCC-based protocol described in the Computational Details section of the main text.

| Solvent[a] | $S$[b] | $\alpha$ | $\mu_0$[c] | $f$[d] | $\mu_1$[e] | $\sigma$[f] | $\mu_1$[g] | $\Delta\mu_{10}$[e] | $\sigma$[f] | $\mu_1/\mu_0$[e] | $\sigma$[f] |
|---|---|---|---|---|---|---|---|---|---|---|---|
| MeOH | 6001 | 30.5 | 4.4 | 0.8546 | 19.6 | 0.2 | 20.0 | 15.2 | 0.2 | 4.4 | 0.1 |
| EtOH | 5809 | 30.5 | 4.3 | 0.8117 | 19.6 | 0.3 | 19.7 | 15.3 | 0.3 | 4.6 | 0.1 |
| $n$-PrOH | 5582 | 30.5 | 4.2 | 0.7802 | 19.5 | 0.3 | 19.6 | 15.3 | 0.3 | 4.6 | 0.1 |
| $i$-PrOH | 5584 | 30.5 | 4.2 | 0.7644 | 19.7 | 0.5 | 19.1 | 15.5 | 0.5 | 4.7 | 0.1 |

[a] Abbreviations: MeOH = methanol, EtOH = ethanol, $n$-PrOH = $n$-propanol, $i$-PrOH = $i$-propanol.
[b] Solvatochromic shift measured from the absorption and fluorescence maxima for **FR0**-SB in each solvent in cm$^{-1}$.
[c] Theoretical values taken from Table III of the main text.
[d] Parameter for each solvent obtained using Eq. (S2).
[e] Calculated using the derived constant $\alpha$, the theoretical values for $\mu_0$ reported in Table III of the main text, and the experimentally measured solvatochromic shift.
[f] Standard deviation of the value reported.
[g] Theoretical values taken from Table IV of the main text.

# Optimized Geometries and Electronic Dipole Moments Resulting from Quantum Chemistry Computations

In what follows, we provide the Cartesian coordinates in XYZ format characterizing the minimum-energy structures of the bare and solvated **FR0**-SB system in the $S_0$ and $S_1$ electronic states, which we used to perform the single-point CC/EOMCC and DFT/TD-DFT calculations described in the main text. For each optimized structure we also provide the Cartesian components of the calculated electronic dipole moment vectors characterizing the $S_0$ and $S_1$ electronic states of the **FR0**-SB chromophore.

**FR0-SB, CAM-B3LYP/6-31+G\* optimized geometry of the $S_0$ state (in Å) along with the electronic dipole moments (in Debye) characterizing the $S_0$ and $S_1$ electronic states, calculated using the CC/EOMCC-based protocol described in the main text**
Resulting from our earlier calculations reported in Ref. 21 of the main text.

|   | X | Y | Z |   | X | Y | Z |
|---|---|---|---|---|---|---|---|
| C | -2.0400612102 | 2.4666616523 | 0.0915028867 | N | -4.8014266006 | 0.1348415258 | 0.6224301494 |
| C | -2.6809013380 | 1.2363881260 | 0.2665672918 | C | -6.2357824232 | 0.2284077493 | 0.7966165612 |
| C | -1.9222545650 | 0.0549395492 | 0.2736457236 | H | -6.5857429379 | 1.2757139840 | 0.7730632763 |
| C | -0.5543425322 | 0.1270187698 | 0.1095601532 | H | -6.4755566946 | -0.1740811216 | 1.7896980632 |
| C | 0.0846040957 | 1.3695023099 | -0.0645565141 | C | -6.9747009202 | -0.5932361483 | -0.2586306514 |
| C | -0.6591570204 | 2.5448550818 | -0.0749668445 | H | -6.5744563404 | -1.6145111705 | -0.2474881934 |
| H | -2.6338579725 | 3.3777809170 | 0.0877126429 | H | -6.7513614399 | -0.1828431295 | -1.2523387551 |
| H | -2.4394533910 | -0.8898411075 | 0.4126501209 | C | -8.4862001069 | -0.6219182137 | -0.0377956745 |
| H | -0.1789196869 | 3.5101908404 | -0.2081431274 | H | -8.8770079440 | 0.4045640463 | -0.0346541686 |
| C | 0.4591709760 | -1.0109545665 | 0.0895034991 | H | -8.6986863427 | -1.0335516438 | 0.9580116942 |
| C | 1.7610097289 | -0.2457806580 | -0.1201205973 | C | -9.2231730528 | -1.4424184123 | -1.0946603084 |
| C | 3.0396730033 | -0.7636035169 | -0.2144075082 | H | -10.3028431011 | -1.4552106085 | -0.9114889679 |
| C | 4.1441127169 | 0.0958809156 | -0.4145993900 | H | -9.0599606925 | -1.0321246604 | -2.0978596187 |
| C | 3.8854930039 | 1.4850749390 | -0.4865354284 | H | -8.8734734786 | -2.4811852408 | -1.1014832350 |
| C | 2.5971124430 | 1.9930128067 | -0.3895445545 | N | 5.4311073075 | -0.4034577000 | -0.5511491106 |
| C | 1.5216408930 | 1.1309095891 | -0.2073078860 | C | 6.5765673612 | 0.4880762006 | -0.6444285372 |
| H | 3.1833474975 | -1.8334882386 | -0.1256243192 | C | 7.0246682547 | 1.1045008737 | 0.6834133829 |
| H | 4.6994735992 | 2.1850753222 | -0.6223288482 | H | 6.3638800367 | 1.2802751212 | -1.3716750441 |
| H | 2.4467626974 | 3.0675911891 | -0.4546427770 | H | 7.4017141243 | -0.0905276937 | -1.0700489463 |
| C | 0.4633037144 | -1.7750518501 | 1.4233308958 | H | 7.8760240181 | 1.7755258017 | 0.5226667709 |
| H | -0.5055588213 | -2.2564475048 | 1.5936213606 | H | 6.2195354359 | 1.6808843982 | 1.1478607110 |
| H | 1.2318582200 | -2.5559008496 | 1.4213128046 | H | 7.3312120781 | 0.3289203345 | 1.3913183989 |
| H | 0.6616623338 | -1.1010938635 | 2.2620728861 | C | 5.6864978806 | -1.8333531841 | -0.4874865096 |
| C | 0.1852892981 | -1.9814899868 | -1.0712141949 | C | 5.7797737551 | -2.4123970352 | 0.9273243308 |
| H | 0.9499166306 | -2.7652570505 | -1.1080497229 | H | 6.6258543682 | -2.0219163119 | -1.0166311879 |
| H | -0.7893333138 | -2.4651977109 | -0.9467832849 | H | 4.9137216851 | -2.3596307378 | -1.0581304232 |
| H | 0.1865968239 | -1.4568843935 | -2.0312793757 | H | 5.9146631764 | -3.4993376762 | 0.8860764646 |
| C | -4.1407804713 | 1.2042617826 | 0.4488338932 | H | 6.6293344403 | -1.9876729820 | 1.4707518587 |
| H | -4.6430337699 | 2.1858299233 | 0.4315704720 | H | 4.8752515706 | -2.2014727821 | 1.5051674126 |

$\mu_0 = 2.580\mathbf{i} – 0.032\mathbf{j} – 0.158\mathbf{k}$,  $\mu_1 = 8.470\mathbf{i} – 1.458\mathbf{j} – 0.706\mathbf{k}$

**FR0-SB, CAM-B3LYP/6-31+G\* optimized geometry of the S$_1$ state (in Å) along with the electronic dipole moments (in Debye) characterizing the S$_0$ and S$_1$ electronic states, calculated using the CC/EOMCC-based protocol described in the main text**

|   | X | Y | Z |   | X | Y | Z |
|---|---|---|---|---|---|---|---|
| C | -2.0113687977 | 2.4966296793 | 0.0542997445 | N | -4.7769845698 | 0.1356746460 | 0.6183269068 |
| C | -2.6817706440 | 1.2550417258 | 0.2565046592 | C | -6.2050385192 | 0.2372494422 | 0.7828928914 |
| C | -1.9174167822 | 0.0467661894 | 0.2897941571 | H | -6.5556433326 | 1.2854307236 | 0.7457713737 |
| C | -0.5615318139 | 0.1091942431 | 0.1274806794 | H | -6.4663027607 | -0.1537988134 | 1.7784041718 |
| C | 0.1067056245 | 1.3603952100 | -0.0775477786 | C | -6.9516819768 | -0.5907972729 | -0.2668571366 |
| C | -0.6514925518 | 2.5716610366 | -0.1110523132 | H | -6.5529888686 | -1.6126636444 | -0.2480337775 |
| H | -2.6084610877 | 3.4054188791 | 0.0323581590 | H | -6.7259402817 | -0.1876076781 | -1.2630577366 |
| H | -2.4495981449 | -0.8857143760 | 0.4474412345 | C | -8.4633399670 | -0.6147770705 | -0.0460577225 |
| H | -0.1614375261 | 3.5285765458 | -0.2655845135 | H | -8.8519930879 | 0.4126699572 | -0.0489724226 |
| C | 0.4485629341 | -1.0339009108 | 0.1328672588 | H | -8.6767273962 | -1.0195251425 | 0.9525387576 |
| C | 1.7523530630 | -0.2784644062 | -0.0960946941 | C | -9.2034990410 | -1.4400286055 | -1.0973281492 |
| C | 3.0178647928 | -0.7794216340 | -0.1921164467 | H | -10.2834222063 | -1.4510057281 | -0.9134121750 |
| C | 4.1273934900 | 0.1010965691 | -0.4122557910 | H | -9.0405219446 | -1.0356433035 | -2.1031320760 |
| C | 3.8623188600 | 1.4988930362 | -0.5017777726 | H | -8.8544184589 | -2.4791759371 | -1.0993484968 |
| C | 2.5868248565 | 2.0107275552 | -0.4046214198 | N | 5.4070471535 | -0.3894440169 | -0.5367565053 |
| C | 1.4912696016 | 1.1282826278 | -0.2072414738 | C | 6.5582631495 | 0.5044158201 | -0.5882809229 |
| H | 3.1783117465 | -1.8452112486 | -0.0841467841 | C | 6.9379615441 | 1.1249427867 | 0.7586200887 |
| H | 4.6788718107 | 2.1918189587 | -0.6589615767 | H | 6.3744411407 | 1.2920591244 | -1.3284216884 |
| H | 2.4261944632 | 3.0814299714 | -0.4821221743 | H | 7.3999875425 | -0.0785429716 | -0.9698898922 |
| C | 0.4535825359 | -1.7664887168 | 1.4845335803 | H | 7.7950843387 | 1.7950340166 | 0.6308031043 |
| H | -0.5229876239 | -2.2260461724 | 1.6692207917 | H | 6.1121109079 | 1.7043293577 | 1.1799735999 |
| H | 1.2111188221 | -2.5589575964 | 1.4974780192 | H | 7.2118409783 | 0.3531582149 | 1.4832531281 |
| H | 0.6630525984 | -1.0750993819 | 2.3059714277 | C | 5.6713248689 | -1.8228233357 | -0.5676808239 |
| C | 0.1706912110 | -2.0295456460 | -1.0053655125 | C | 5.8579554956 | -2.4650731518 | 0.8092433390 |
| H | 0.9258882230 | -2.8239295073 | -1.0227220216 | H | 6.5750587047 | -1.9741112470 | -1.1663620213 |
| H | -0.8107034510 | -2.4956937761 | -0.8694167444 | H | 4.8628993704 | -2.3193329602 | -1.1109603972 |
| H | 0.1757873407 | -1.5272467014 | -1.9770698843 | H | 6.0166420172 | -3.5436373222 | 0.6988899465 |
| C | -4.1109786006 | 1.2255639791 | 0.4294321650 | H | 6.7269740956 | -2.0476461912 | 1.3263931132 |
| H | -4.6258039512 | 2.1994602474 | 0.3965937069 | H | 4.9804771052 | -2.3097200770 | 1.4437928409 |

$\mu_0$ = 3.402**i** – 0.117**j** – 0.267**k**, $\mu_1$ = 10.755**i** – 1.419**j** – 0.945**k**

**FR0-SB in methanol, CAM-B3LYP/6-31+G*/SMD optimized geometry of the $S_0$ state (in Å) along with the electronic dipole moments (in Debye) characterizing the $S_0$ and $S_1$ electronic states of the FR0-SB chromophore, calculated using the CC/EOMCC-based protocol described in the main text**

The curly brackets indicate the three alcohol solvent molecules being replaced by effective fragment potentials (EFPs) in the single-point DFT/TD-DFT and CC/EOMCC calculations described in the main text.

|   | X | Y | Z |   | X | Y | Z |   |
|---|---|---|---|---|---|---|---|---|
| C | -0.0574152725 | -2.3420067124 | 1.6815631788 | H | 6.9267340249 | -0.9925823566 | -0.5512228484 | |
| C | 0.6506680444 | -1.4846548092 | 0.8285402574 | C | 7.1669831043 | -3.0203533299 | -1.2457016815 | |
| C | -0.0543221337 | -0.6808998671 | -0.0861080704 | H | 8.2572016521 | -3.0177599109 | -1.1343482446 | |
| C | -1.4329550338 | -0.7439359592 | -0.1224976535 | H | 6.8125461444 | -4.0328213384 | -1.0169226229 | |
| C | -2.1344429806 | -1.6121959019 | 0.7384537647 | H | 6.9405843542 | -2.8191023485 | -2.2998478659 | |
| C | -1.4465004503 | -2.4156079970 | 1.6437481738 | N | -7.4264333841 | -0.4152160297 | -0.8253600533 | |
| H | 0.4910370289 | -2.9596580475 | 2.3890035051 | C | -8.6096865021 | -0.9482968013 | -0.1575657400 | |
| H | 0.4923797232 | -0.0098524870 | -0.7417669459 | C | -8.9523915396 | -0.2857233286 | 1.1765206566 | |
| H | -1.9760922294 | -3.0868095518 | 2.3143528817 | H | -8.4960692714 | -2.0300893526 | -0.0223408260 | |
| C | -2.4000211434 | 0.0418782869 | -1.0017942275 | H | -9.4508167363 | -0.8261984594 | -0.8453241853 | |
| C | -3.7433980356 | -0.4959438033 | -0.5196970489 | H | -9.8207294278 | -0.7817189217 | 1.6258292584 | |
| C | -5.0060585120 | -0.1269715129 | -0.9476421264 | H | -8.1236528869 | -0.3483449270 | 1.8884131950 | |
| C | -6.1569287061 | -0.7335632109 | -0.3861076413 | H | -9.2002858489 | 0.7719311841 | 1.0422513863 | |
| C | -5.9551796020 | -1.6952483568 | 0.6375517291 | C | -7.6300828506 | 0.5401346938 | -1.9079455625 | |
| C | -4.6830580851 | -2.0575184463 | 1.0573544128 | C | -7.6316235299 | 2.0090383991 | -1.4853631391 | |
| C | -3.5648983327 | -1.4572319753 | 0.4840792354 | H | -8.5879537612 | 0.3008705379 | -2.3800251896 | |
| H | -5.1048401800 | 0.6346972133 | -1.7112432737 | H | -6.8737940823 | 0.3699594884 | -2.6812832737 | |
| H | -6.8017141662 | -2.1760543092 | 1.1104177456 | H | -7.7363127505 | 2.6518283927 | -2.3673826273 | |
| H | -4.5801273310 | -2.8033258850 | 1.8421221631 | H | -8.4670008989 | 2.2264965824 | -0.8122763148 | |
| C | -2.2736708650 | 1.5535471007 | -0.7513459875 | H | -6.7054019714 | 2.2870637783 | -0.9722991136 | |
| H | -1.2877592194 | 1.9167218001 | -1.0598291516 | C | 3.1691727998 | 0.6224330720 | -2.9795646648 | ⎫ |
| H | -3.0254866926 | 2.1052117778 | -1.3270122826 | O | 2.8464167397 | 1.1921378697 | -1.7080725087 | |
| H | -2.4172245380 | 1.7932855185 | 0.3076674287 | H | 3.1296610664 | 1.4218138977 | -3.7227189201 | ⎬ EFP |
| C | -2.1775912168 | -0.2650970755 | -2.4915483889 | H | 2.4490238645 | -0.1552492991 | -3.2543679887 | |
| H | -2.9108614549 | 0.2649241201 | -3.1093293723 | H | 4.1795925579 | 0.1999539905 | -2.9768002929 | |
| H | -1.1802910299 | 0.0568460435 | -2.8087391514 | H | 2.7586527173 | 0.4610676601 | -1.0294997020 | ⎭ |
| H | -2.2720598950 | -1.3365232085 | -2.6973505934 | C | 4.6989308932 | 4.1561643900 | -1.3362125340 | ⎫ |
| C | 2.1116816138 | -1.4648075926 | 0.9473154269 | O | 4.9584700930 | 2.8177920513 | -0.9298783395 | |
| H | 2.5147897493 | -2.0887076109 | 1.7567599989 | H | 5.6285800480 | 4.7209988015 | -1.2289266856 | ⎬ EFP |
| N | 2.9127670136 | -0.7997691554 | 0.2099603162 | H | 3.9345311835 | 4.6253317541 | -0.7039548846 | |
| C | 4.3420728436 | -0.9460511991 | 0.4786777382 | H | 4.3812508708 | 4.2064767656 | -2.3851012940 | |
| H | 4.5142874398 | -1.2608388802 | 1.5177707694 | H | 4.1870947963 | 2.2563701459 | -1.1676824937 | ⎭ |
| H | 4.8087036996 | 0.0379432952 | 0.3586576040 | C | 0.7966935287 | 4.0801944490 | -1.2675820013 | ⎫ |
| C | 4.9941966417 | -1.9437421822 | -0.4784989567 | O | 0.7796441804 | 2.9878787822 | -2.1754750584 | |
| H | 4.7289163970 | -1.6837367512 | -1.5113376383 | H | -0.1646877571 | 4.5950457122 | -1.3453458496 | ⎬ EFP |
| H | 4.5719027209 | -2.9405014396 | -0.2918212702 | H | 1.5900506730 | 4.7971703529 | -1.5152462062 | |
| C | 6.5145050383 | -1.9864358941 | -0.3317481252 | H | 0.9280240108 | 3.7450895814 | -0.2312886331 | |
| H | 6.7756835011 | -2.2064038177 | 0.7120382931 | H | 1.5010521768 | 2.3622918549 | -1.9490446976 | ⎭ |

$\boldsymbol{\mu}_0 = -4.292\mathbf{i} - 0.457\mathbf{j} + 0.553\mathbf{k}, \quad \boldsymbol{\mu}_1 = -16.234\mathbf{i} + 1.037\mathbf{j} - 1.680\mathbf{k}$

**FR0-SB in methanol, CAM-B3LYP/6-31+G*/SMD optimized geometry of the $S_1$ state (in Å) along with the electronic dipole moments (in Debye) characterizing the $S_0$ and $S_1$ electronic states of the FR0-SB chromophore, calculated using the CC/EOMCC-based protocol described in the main text**

The curly brackets indicate the three alcohol solvent molecules being replaced by effective fragment potentials (EFPs) in the single-point DFT/TD-DFT and CC/EOMCC calculations described in the main text.

|   | X | Y | Z |   | X | Y | Z |   |
|---|---|---|---|---|---|---|---|---|
| C | -0.0442551982 | -2.3607008381 | 1.6082181330 | H | 6.9424427062 | -1.1708366049 | -0.5312864782 | |
| C | 0.6886177823 | -1.4670205693 | 0.7567832133 | C | 7.0684488305 | -3.2481001925 | -1.0959000413 | |
| C | -0.0465092804 | -0.6307753067 | -0.1546648509 | H | 8.1535220303 | -3.3083877059 | -0.9524192173 | |
| C | -1.4107946746 | -0.6934156182 | -0.1767555351 | H | 6.6427963799 | -4.2205769520 | -0.8197221653 | |
| C | -2.1316277123 | -1.5856805193 | 0.6881976476 | H | 6.8834523531 | -3.0966232847 | -2.1663914963 | |
| C | -1.4091993026 | -2.4292742328 | 1.5937256101 | N | -7.3879762059 | -0.4394182722 | -0.7904740708 | |
| H | 0.5175563653 | -2.9958171112 | 2.2898463764 | C | -8.5675618180 | -0.9871722466 | -0.1121951993 | |
| H | 0.4980583338 | 0.0450199047 | -0.8064206107 | C | -8.8715705560 | -0.3411124869 | 1.2385840286 | |
| H | -1.9363078212 | -3.1132740503 | 2.2531102036 | H | -8.4485906045 | -2.0700134658 | -0.0034997286 | |
| C | -2.3881340013 | 0.0954719090 | -1.0448774783 | H | -9.4171352122 | -0.8464143534 | -0.7832470567 | |
| C | -3.7253489076 | -0.4519132645 | -0.5585587222 | H | -9.7362953652 | -0.8392846486 | 1.6906531352 | |
| C | -4.9785250771 | -0.1120606722 | -0.9664595185 | H | -8.0312732384 | -0.4261997228 | 1.9336172496 | |
| C | -6.1287211447 | -0.7484158746 | -0.3811884265 | H | -9.1120844105 | 0.7202660430 | 1.1271398630 | |
| C | -5.9033770685 | -1.7259336001 | 0.6426121541 | C | -7.6274508974 | 0.4899750049 | -1.8977580347 | |
| C | -4.6396555524 | -2.0725496768 | 1.0513510964 | C | -7.6760101426 | 1.9598823355 | -1.4840324212 | |
| C | -3.5093549985 | -1.4405658684 | 0.4615950287 | H | -8.5776663928 | 0.2069541570 | -2.3580409908 | |
| H | -5.1057487672 | 0.6493279052 | -1.7254501578 | H | -6.8672229616 | 0.3293814284 | -2.6662706308 | |
| H | -6.7442208596 | -2.2249758043 | 1.1062387028 | H | -7.8051742338 | 2.5842272956 | -2.3751295691 | |
| H | -4.5125511415 | -2.8257611557 | 1.8233076060 | H | -8.5162674618 | 2.1584496173 | -0.8122558335 | |
| C | -2.2687874422 | 1.6060334789 | -0.7877543192 | H | -6.7564219895 | 2.2720995008 | -0.9793394361 | |
| H | -1.2812698512 | 1.9693442008 | -1.0906445142 | C | 3.1504352420 | 0.7110695853 | -2.9263006742 | ⎫ |
| H | -3.0192787141 | 2.1603589075 | -1.3629590228 | O | 2.8354038108 | 1.2836172622 | -1.6581178434 | ⎬ EFP |
| H | -2.4140553283 | 1.8407152608 | 0.2719902254 | H | 3.1737260631 | 1.5152160412 | -3.6662268549 | |
| C | -2.1791134273 | -0.2058053340 | -2.5377791441 | H | 2.3935824347 | -0.0228531904 | -3.2252156736 | |
| H | -2.9110567961 | 0.3325716212 | -3.1501954006 | H | 4.1335115444 | 0.2269867776 | -2.9092423209 | |
| H | -1.1799387033 | 0.1088208231 | -2.8552297480 | H | 2.7540862249 | 0.5379784323 | -0.9664638409 | ⎭ |
| H | -2.2818768432 | -1.2755716269 | -2.7475392478 | C | 4.6758650938 | 4.2055007169 | -1.3777144914 | ⎫ |
| C | 2.1096094365 | -1.4492736372 | 0.8507274099 | O | 4.9704565150 | 2.8754399093 | -0.9715641785 | ⎬ EFP |
| H | 2.5402999919 | -2.1398733326 | 1.5876531721 | H | 5.5966986210 | 4.7894232981 | -1.2981174184 | |
| N | 2.9207622174 | -0.6831133728 | 0.1639473044 | H | 3.9189642827 | 4.6644932201 | -0.7290439031 | |
| C | 4.3414626612 | -0.8780936822 | 0.4112505403 | H | 4.3297807640 | 4.2443961882 | -2.4183236203 | |
| H | 4.5269706312 | -1.1075640459 | 1.4729474411 | H | 4.1907029215 | 2.3049184909 | -1.1667369868 | ⎭ |
| H | 4.8599647939 | 0.0641578224 | 0.1953476167 | C | 0.7796616565 | 4.1752651167 | -1.2468073002 | ⎫ |
| C | 4.9502299708 | -1.9905591622 | -0.4508497964 | O | 0.8112902962 | 3.0956724234 | -2.1677556297 | ⎬ EFP |
| H | 4.7209164129 | -1.7923469021 | -1.5061866698 | H | -0.1904044343 | 4.6711725285 | -1.3422971128 | |
| H | 4.4598127961 | -2.9409250699 | -0.1990871586 | H | 1.5632236251 | 4.9132888307 | -1.4627219950 | |
| C | 6.4610139185 | -2.1212657276 | -0.2640945950 | H | 0.8912776919 | 3.8296572054 | -0.2116073405 | |
| H | 6.6841501708 | -2.2920380460 | 0.7978137679 | H | 1.5247225981 | 2.4674432862 | -1.9160838846 | ⎭ |

$\mu_0 = -6.582\mathbf{i} - 0.712\mathbf{j} + 0.168\mathbf{k}$,   $\mu_1 = -19.883\mathbf{i} + 0.429\mathbf{j} - 1.634\mathbf{k}$

**FR0-SB in ethanol, CAM-B3LYP/6-31+G*/SMD optimized geometry of the $S_0$ state (in Å) along with the electronic dipole moments (in Debye) characterizing the $S_0$ and $S_1$ electronic states of the FR0-SB chromophore, calculated using the CC/EOMCC-based protocol described in the main text**

The curly brackets indicate the three alcohol solvent molecules being replaced by effective fragment potentials (EFPs) in the single-point DFT/TD-DFT and CC/EOMCC calculations described in the main text.

|   | X | Y | Z |   | X | Y | Z |   |
|---|---|---|---|---|---|---|---|---|
| C | -0.0729117803 | -2.3272442308 | 1.6290391343 | N | -7.4889943900 | -0.4805408102 | -0.7988892705 | |
| C | 0.6165543207 | -1.4506234963 | 0.7802224853 | C | -8.6595760873 | -1.0304560555 | -0.1229417588 | |
| C | -0.1070525809 | -0.6502273464 | -0.1227905701 | C | -8.9965797422 | -0.3767986276 | 1.2170748960 | |
| C | -1.4852176667 | -0.7366753956 | -0.1526583914 | H | -8.5320773625 | -2.1113174906 | 0.0070145094 | |
| C | -2.1670306494 | -1.6244462901 | 0.7042789654 | H | -9.5086275536 | -0.9153396683 | -0.8021423728 | |
| C | -1.4605138361 | -2.4247369908 | 1.5978945310 | H | -9.8539431888 | -0.8846144833 | 1.6741988256 | |
| H | 0.4898194976 | -2.9409727300 | 2.3286620633 | H | -8.1593307699 | -0.4317036502 | 1.9195809612 | |
| H | 0.4240704173 | 0.0407687083 | -0.7705022448 | H | -9.2584714473 | 0.6782400364 | 1.0887457511 | |
| H | -1.9744909543 | -3.1106512060 | 2.2658665609 | C | -7.7125577242 | 0.4734883942 | -1.8786020064 | |
| C | -2.4717183803 | 0.0429814035 | -1.0161643083 | C | -7.7229454374 | 1.9422248087 | -1.4549404502 | |
| C | -3.8023798304 | -0.5215081424 | -0.5306044577 | H | -8.6729819463 | 0.2261782100 | -2.3412889063 | |
| C | -5.0735610276 | -0.1651547564 | -0.9436861200 | H | -6.9622939340 | 0.3109148108 | -2.6594600317 | |
| C | -6.2115721882 | -0.7890038751 | -0.3752648596 | H | -7.8385288755 | 2.5848341095 | -2.3357423110 | |
| C | -5.9881404780 | -1.7555820849 | 0.6392445202 | H | -8.5552829978 | 2.1527883982 | -0.7759321577 | |
| C | -4.7074024723 | -2.1058772205 | 1.0433084937 | H | -6.7954581125 | 2.2270433211 | -0.9480099410 | |
| C | -3.6020601481 | -1.4885400882 | 0.4632474217 | C | 3.4196540045 | 1.7034301417 | -4.1200113486 | ⎫ |
| H | -5.1882739394 | 0.6021689132 | -1.6993167085 | C | 3.0267498183 | 0.6376903630 | -3.1205941448 | ⎪ |
| H | -6.8238314775 | -2.2493893896 | 1.1181196289 | O | 2.8828171096 | 1.2304393996 | -1.8202573242 | ⎪ |
| H | -4.5880208920 | -2.8558987323 | 1.8217173282 | H | 3.4891283627 | 1.2645127223 | -5.1213069473 | ⎪ |
| C | -2.3688446660 | 1.5533799689 | -0.7487451151 | H | 2.6771410128 | 2.5077559819 | -4.1542285173 | ⎬ EFP |
| H | -1.3892228462 | 1.9367366150 | -1.0511381211 | H | 4.3950670199 | 2.1362201745 | -3.8735984425 | ⎪ |
| H | -3.1318164917 | 2.0988890698 | -1.3158283664 | H | 2.0793173776 | 0.1663203904 | -3.4094585308 | ⎪ |
| H | -2.5152370446 | 1.7776911026 | 0.3135318298 | H | 3.7927219571 | -0.1449913071 | -3.0772863740 | ⎪ |
| C | -2.2563100783 | -0.2420918050 | -2.5113241470 | H | 2.7537459269 | 0.5106871918 | -1.1403465118 | ⎭ |
| H | -3.0092405627 | 0.2756654339 | -3.1157637521 | C | 6.3360353876 | 4.7578250224 | -0.9503991425 | ⎫ |
| H | -1.2713093109 | 0.1099382465 | -2.8351514722 | C | 5.0167893518 | 4.0864308208 | -1.2660269936 | ⎪ |
| H | -2.3250886892 | -1.3131678409 | -2.7285556717 | O | 5.1000570083 | 2.7067342083 | -0.9148961435 | ⎪ |
| C | 2.0775978345 | -1.4049550096 | 0.8997607503 | H | 6.2889588019 | 5.8226279324 | -1.2051355771 | ⎪ |
| H | 2.4859118220 | -2.0071405020 | 1.7231988437 | H | 6.5723962939 | 4.6745275339 | 0.1160438720 | ⎬ EFP |
| N | 2.8738427819 | -0.7440115947 | 0.1537692112 | H | 7.1531767181 | 4.3052865889 | -1.5225716413 | ⎪ |
| C | 4.3019098801 | -0.8615566649 | 0.4465834726 | H | 4.2065327783 | 4.5686240498 | -0.7024344803 | ⎪ |
| H | 4.4617010856 | -1.1471913785 | 1.4960808397 | H | 4.7879949798 | 4.1900559553 | -2.3354739413 | ⎪ |
| H | 4.7556446350 | 0.1253183872 | 0.3062452007 | H | 4.3043748145 | 2.2325599176 | -1.2412585042 | ⎭ |
| C | 4.9872747374 | -1.8740397822 | -0.4706078123 | C | -0.1284968279 | 5.1810149579 | -1.3256073052 | ⎫ |
| H | 4.7630915242 | -1.6289653932 | -1.5165922635 | C | 1.0916182665 | 4.2844423868 | -1.3360534584 | ⎪ |
| H | 4.5583762206 | -2.8681043779 | -0.2847962771 | O | 0.7160317396 | 2.9952919474 | -1.8092057222 | ⎪ |
| C | 6.5010390427 | -1.9147799122 | -0.2665586351 | H | 0.1335388300 | 6.1749523848 | -0.9455376065 | ⎪ |
| H | 6.7227408515 | -2.1202363185 | 0.7892266464 | H | -0.5358447986 | 5.2996853112 | -2.3358359647 | ⎬ EFP |
| H | 6.9223533483 | -0.9243520868 | -0.4838343723 | H | -0.9120321373 | 4.7638404868 | -0.6839649976 | ⎪ |
| C | 7.1860888454 | -2.9618731720 | -1.1407575972 | H | 1.8674596731 | 4.7150732268 | -1.9846441912 | ⎪ |
| H | 8.2715696664 | -2.9573201487 | -0.9900719582 | H | 1.5074548132 | 4.2061759763 | -0.3216042858 | ⎪ |
| H | 6.8230502057 | -3.9709378615 | -0.9103023426 | H | 1.4988151874 | 2.4031363639 | -1.7930097891 | ⎭ |
| H | 6.9983439535 | -2.7763908173 | -2.2053176361 |   |   |   |   | |

$\boldsymbol{\mu}_0 = -4.201\mathbf{i} - 0.327\mathbf{j} + 0.666\mathbf{k}, \quad \boldsymbol{\mu}_1 = -15.840\mathbf{i} + 1.022\mathbf{j} - 1.436\mathbf{k}$

**FR0-SB in ethanol, CAM-B3LYP/6-31+G*/SMD optimized geometry of the S$_1$ state (in Å) along with the electronic dipole moments (in Debye) characterizing the S$_0$ and S$_1$ electronic states of the FR0-SB chromophore, calculated using the CC/EOMCC-based protocol described in the main text**

The curly brackets indicate the three alcohol solvent molecules being replaced by effective fragment potentials (EFPs) in the single-point DFT/TD-DFT and CC/EOMCC calculations described in the main text.

|   | X | Y | Z |   | X | Y | Z |   |
|---|---|---|---|---|---|---|---|---|
| C | -0.0580542632 | -2.3755930017 | 1.5404437977 | N | -7.4430073687 | -0.4973581696 | -0.7653646734 | |
| C | 0.6581050160 | -1.4577239529 | 0.7005325328 | C | -8.6115523915 | -1.0630748347 | -0.0834786925 | |
| C | -0.0941909756 | -0.6129522955 | -0.1888883189 | C | -8.9106404405 | -0.4310138795 | 1.2750375899 | |
| C | -1.4580747765 | -0.6925657230 | -0.2031678710 | H | -8.4807796807 | -2.1454418380 | 0.0153742168 | |
| C | -2.1614511918 | -1.6110710721 | 0.6491124169 | H | -9.4678387324 | -0.9249046866 | -0.7466260071 | |
| C | -1.4219991733 | -2.4622164325 | 1.5335107695 | H | -9.7668298723 | -0.9406841557 | 1.7305933024 | |
| H | 0.5163192697 | -3.0155614261 | 2.2069389283 | H | -8.0635957747 | -0.5137258667 | 1.9622488804 | |
| H | 0.4367710636 | 0.0870108870 | -0.8261404183 | H | -9.1619210225 | 0.6289542322 | 1.1740518767 | |
| H | -1.9352689354 | -3.1653521521 | 2.1836365982 | C | -7.7001028406 | 0.4390539052 | -1.8624625732 | |
| C | -2.4527287394 | 0.1021455041 | -1.0462796431 | C | -7.7636666278 | 1.9046822214 | -1.4357507714 | |
| C | -3.7785732396 | -0.4734742446 | -0.5625179949 | H | -8.6495418391 | 0.1487602319 | -2.3198597402 | |
| C | -5.0388446681 | -0.1411549103 | -0.9545630505 | H | -6.9420992229 | 0.2946346017 | -2.6362681615 | |
| C | -6.1770118809 | -0.7998374303 | -0.3708150483 | H | -7.9039048463 | 2.5354120554 | -2.3207336865 | |
| C | -5.9325257113 | -1.7902812317 | 0.6358666379 | H | -8.6027589434 | 2.0885024971 | -0.7583810532 | |
| C | -4.6613047682 | -2.1310589531 | 1.0267933470 | H | -6.8452785463 | 2.2228434062 | -0.9327292319 | |
| C | -3.5428038003 | -1.4780722195 | 0.4371550337 | C | 3.4183464535 | 1.7495053068 | -4.0751684204 | ⎤ |
| H | -5.1803198022 | 0.6325627099 | -1.6984256251 | C | 3.0193966330 | 0.7010991615 | -3.0582402634 | ⎥ |
| H | -6.7638653192 | -2.3040837972 | 1.1006611088 | O | 2.8530337689 | 1.3081095309 | -1.7711777129 | ⎥ |
| H | -4.5192431466 | -2.8955965818 | 1.7848858378 | H | 3.4999802514 | 1.2923508079 | -5.0676614515 | ⎥ |
| C | -2.3525323185 | 1.6080778987 | -0.7568540687 | H | 2.6727991435 | 2.5497371946 | -4.1318863997 | ⎬ EFP |
| H | -1.3722082297 | 1.9928520096 | -1.0552881592 | H | 4.3889899211 | 2.1921045339 | -3.8273707955 | ⎥ |
| H | -3.1154492838 | 2.1629015162 | -1.3153954312 | H | 2.0801183828 | 0.2168500534 | -3.3550033076 | ⎥ |
| H | -2.4967391735 | 1.8172653246 | 0.3086853428 | H | 3.7899121796 | -0.0764839410 | -2.9937983652 | ⎥ |
| C | -2.2514079873 | -0.1619840369 | -2.5472369486 | H | 2.7435054102 | 0.5788457603 | -1.0742185190 | ⎦ |
| H | -3.0014472488 | 0.3727769420 | -3.1407446108 | C | 6.2981468795 | 4.8290825402 | -1.0102512847 | ⎤ |
| H | -1.2632526018 | 0.1837404626 | -2.8674517200 | C | 4.9724060168 | 4.1473080593 | -1.2731605431 | ⎥ |
| H | -2.3314736498 | -1.2289536531 | -2.7795329910 | O | 5.0742727487 | 2.7720953804 | -0.9133796087 | ⎥ |
| C | 2.0795580527 | -1.4225580668 | 0.7918070257 | H | 6.2370079075 | 5.8911179723 | -1.2737390053 | ⎥ |
| H | 2.5158457863 | -2.1074223207 | 1.5313266718 | H | 6.5721461107 | 4.7574847062 | 0.0479792421 | ⎬ EFP |
| N | 2.8846142583 | -0.6492246635 | 0.1062827524 | H | 7.0972050951 | 4.3746359479 | -1.6060335802 | ⎥ |
| C | 4.3047177180 | -0.8189001736 | 0.3808695997 | H | 4.1799565084 | 4.6315022909 | -0.6861644696 | ⎥ |
| H | 4.4746997367 | -1.0115713289 | 1.4527250170 | H | 4.7053969664 | 4.2404583380 | -2.3347293131 | ⎥ |
| H | 4.8146081730 | 0.1220409740 | 0.1426801767 | H | 4.2728542817 | 2.2894672821 | -1.2188841163 | ⎦ |
| C | 4.9461378801 | -1.9505505446 | -0.4314974262 | C | -0.1393622543 | 5.2616887944 | -1.3673352095 | ⎤ |
| H | 4.7548745192 | -1.7801592587 | -1.4991658210 | C | 1.0749615845 | 4.3569196640 | -1.3558875710 | ⎥ |
| H | 4.4521785576 | -2.8973448564 | -0.1730944162 | O | 0.6999556752 | 3.0695371266 | -1.8313781813 | ⎥ |
| C | 6.4499967719 | -2.0688131030 | -0.1890013849 | H | 0.1226337894 | 6.2550933802 | -0.9857091030 | ⎥ |
| H | 6.6358539154 | -2.2136491644 | 0.8839212824 | H | -0.5299548090 | 5.3807438597 | -2.3841533832 | ⎬ EFP |
| H | 6.9365981801 | -1.1224182685 | -0.4610149143 | H | -0.9357756668 | 4.8512231505 | -0.7371868961 | ⎥ |
| C | 7.0911023509 | -3.2119471002 | -0.9718870383 | H | 1.8639388374 | 4.7824590308 | -1.9919432221 | ⎥ |
| H | 8.1702574689 | -3.2652573902 | -0.7874296314 | H | 1.4740218155 | 4.2792419045 | -0.3345797513 | ⎥ |
| H | 6.6580238065 | -4.1793542635 | -0.6892681883 | H | 1.4763988831 | 2.4675040099 | -1.7874666760 | ⎦ |
| H | 6.9450205754 | -3.0858095371 | -2.0516710748 | | | | | |

$\boldsymbol{\mu_0} = -6.463\mathbf{i} - 0.603\mathbf{j} + 0.284\mathbf{k}$,  $\boldsymbol{\mu_1} = -19.622\mathbf{i} + 0.420\mathbf{j} - 1.395\mathbf{k}$

**FR0-SB in *n*-propanol, CAM-B3LYP/6-31+G*/SMD optimized geometry of the S$_0$ state (in Å) along with the electronic dipole moments (in Debye) characterizing the S$_0$ and S$_1$ electronic states of the FR0-SB chromophore, calculated using the CC/EOMCC-based protocol described in the main text**

The curly brackets indicate the three alcohol solvent molecules being replaced by effective fragment potentials (EFPs) in the single-point DFT/TD-DFT and CC/EOMCC calculations described in the main text.

|   | X | Y | Z |   | X | Y | Z |   |
|---|---|---|---|---|---|---|---|---|
| C | -0.0690226790 | -2.3122671680 | 1.6301606704 | H | -9.5040887506 | -0.9062219831 | -0.8067463918 | |
| C | 0.6207687767 | -1.4368616386 | 0.7802318540 | H | -9.8481460412 | -0.8721764852 | 1.6698259944 | |
| C | -0.1024028570 | -0.6398032630 | -0.1261478064 | H | -8.1532418766 | -0.4198875669 | 1.9137460726 | |
| C | -1.4804810603 | -0.7280931447 | -0.1578790879 | H | -9.2520408982 | 0.6896933407 | 1.0820625612 | |
| C | -2.1625217391 | -1.6142086541 | 0.7006368383 | C | -7.7075807979 | 0.4813037992 | -1.8848393871 | |
| C | -1.4564018842 | -2.4113289000 | 1.5973571435 | C | -7.7150673874 | 1.9502804407 | -1.4619286299 | |
| H | 0.4933579322 | -2.9235054805 | 2.3322011115 | H | -8.6690804380 | 0.2354001308 | -2.3460103359 | |
| H | 0.4285993863 | 0.0507263025 | -0.7744873521 | H | -6.9586231083 | 0.3171297799 | -2.6666011184 | |
| H | -1.9705232656 | -3.0959028518 | 2.2666041286 | H | -7.8302233293 | 2.5926206157 | -2.3429823735 | |
| C | -2.4668437005 | 0.0479484252 | -1.0249684447 | H | -8.5461034646 | 2.1626965123 | -0.7819323960 | |
| C | -3.7975544381 | -0.5149203593 | -0.5378037669 | H | -6.7865935200 | 2.2337309189 | -0.9560240011 | |
| C | -5.0686621850 | -0.1590383478 | -0.9514807687 | C | 3.5088093913 | 1.0918775293 | -5.5506623056 | ⎫ |
| C | -6.2067263235 | -0.7811623428 | -0.3814361221 | C | 3.4111513562 | 1.6776708934 | -4.1442776126 | ⎪ |
| C | -5.9835196690 | -1.7460339726 | 0.6346956440 | C | 3.0261441832 | 0.6254287913 | -3.1215673576 | ⎪ |
| C | -4.7028608164 | -2.0958077477 | 1.0394716259 | O | 2.8966753965 | 1.2312072883 | -1.8271669811 | ⎪ |
| C | -3.5974616288 | -1.4796511193 | 0.4582932744 | H | 3.7885707549 | 1.8626301057 | -6.2762324142 | ⎪ |
| H | -5.1832202943 | 0.6069647830 | -1.7084719237 | H | 2.5518716286 | 0.6653526787 | -5.8738708791 | ⎪ |
| H | -6.8193928729 | -2.2388316703 | 1.1142584913 | H | 4.2636219330 | 0.2982054451 | -5.6034793426 | ⎬ EFP |
| H | -4.5836864916 | -2.8441666008 | 1.8194933498 | H | 2.6683152047 | 2.4845076565 | -4.1271331270 | ⎪ |
| C | -2.3641645013 | 1.5595211395 | -0.7642171505 | H | 4.3735195959 | 2.1195555929 | -3.8581951660 | ⎪ |
| H | -1.3844947901 | 1.9411849936 | -1.0681850602 | H | 2.0746499027 | 0.1519659430 | -3.3966645368 | ⎪ |
| H | -3.1269674119 | 2.1024520883 | -1.3340483636 | H | 3.7914224565 | -0.1591849990 | -3.0789163366 | ⎪ |
| H | -2.5109699489 | 1.7886315365 | 0.2969607593 | H | 2.7668810508 | 0.5180888190 | -1.1402748803 | ⎭ |
| C | -2.2508926097 | -0.2434658776 | -2.5187956809 | C | 6.2789058042 | 6.2861464942 | -1.3116299928 | ⎫ |
| H | -3.0038337643 | 0.2715911384 | -3.1255630596 | C | 6.3187357504 | 4.7996415433 | -0.9663157089 | ⎪ |
| H | -1.2659403365 | 0.1075543363 | -2.8437611790 | C | 5.0114615056 | 4.1014550515 | -1.2917130886 | ⎪ |
| H | -2.3193480827 | -1.3153826791 | -2.7318259057 | O | 5.1077466529 | 2.7253782640 | -0.9338612043 | ⎪ |
| C | 2.0814966770 | -1.3891450732 | 0.9031564699 | H | 7.2296660283 | 6.7717123789 | -1.0666090048 | ⎪ |
| H | 2.4881685171 | -1.9898303354 | 1.7283956337 | H | 5.4898208332 | 6.8064221031 | -0.7555638925 | ⎪ |
| N | 2.8795484801 | -0.7288818776 | 0.1584571675 | H | 6.0931987752 | 6.4451948760 | -2.3807272494 | ⎬ EFP |
| C | 4.3068255050 | -0.8449833622 | 0.4557532920 | H | 6.5327395888 | 4.6707690072 | 0.1019012209 | ⎪ |
| H | 4.4635708107 | -1.1305768220 | 1.5058153365 | H | 7.1304732914 | 4.3083625071 | -1.5168175000 | ⎪ |
| H | 4.7598931905 | 0.1425089006 | 0.3167950322 | H | 4.1886546885 | 4.5769765608 | -0.7392929171 | ⎪ |
| C | 4.9963538147 | -1.8566004899 | -0.4594927866 | H | 4.7928329386 | 4.1987803605 | -2.3647926649 | ⎪ |
| H | 4.7746569435 | -1.6119265480 | -1.5060999648 | H | 4.3174578510 | 2.2424164940 | -1.2599649836 | ⎭ |
| H | 4.5685004929 | -2.8514668610 | -0.2748885172 | C | 0.1563320340 | 6.5609453306 | -0.7736193492 | ⎫ |
| C | 6.5096042961 | -1.8950701427 | -0.2514987252 | C | -0.1519580598 | 5.1559206719 | -1.2845976581 | ⎪ |
| H | 6.7287830200 | -2.1009901515 | 0.8047840483 | C | 1.0836654910 | 4.2756734916 | -1.3293433263 | ⎪ |
| H | 6.9297720029 | -0.9037194968 | -0.4670895861 | O | 0.7207663571 | 2.9812346203 | -1.7948675337 | ⎪ |
| C | 7.1986194752 | -2.9404836469 | -1.1247347424 | H | -0.7465913944 | 7.1807966855 | -0.7599268811 | ⎪ |
| H | 8.2836716078 | -2.9345197380 | -0.9712925426 | H | 0.8935483377 | 7.0663473896 | -1.4092197733 | ⎪ |
| H | 6.8366974064 | -3.9504459430 | -0.8959265611 | H | 0.5564017532 | 6.5377207174 | 0.2472168585 | ⎬ EFP |
| H | 7.0131399930 | -2.7546198173 | -2.1896673467 | H | -0.5839913103 | 5.2117948504 | -2.2916557871 | ⎪ |
| N | -7.4841962789 | -0.4724623840 | -0.8049475892 | H | -0.9019359059 | 4.6779473419 | -0.6423553799 | ⎪ |
| C | -8.6547388717 | -1.0207713477 | -0.1278072211 | H | 1.8349717775 | 4.7229626811 | -1.9969071272 | ⎪ |
| C | -8.9908070240 | -0.3653605969 | 1.2115712616 | H | 1.5272878121 | 4.2053111705 | -0.3251904860 | ⎪ |
| H | -8.5277217785 | -2.1015464902 | 0.0034135044 | H | 1.5116561077 | 2.3999075920 | -1.7891902332 | ⎭ |

$\mu_0 = -4.186\mathbf{i} - 0.310\mathbf{j} + 0.663\mathbf{k}$, $\mu_1 = -15.756\mathbf{i} + 1.025\mathbf{j} - 1.441\mathbf{k}$

**FR0-SB in *n*-propanol, CAM-B3LYP/6-31+G*/SMD optimized geometry of the $S_1$ state (in Å) along with the electronic dipole moments (in Debye) characterizing the $S_0$ and $S_1$ electronic states of the FR0-SB chromophore, calculated using the CC/EOMCC-based protocol described in the main text**

The curly brackets indicate the three alcohol solvent molecules being replaced by effective fragment potentials (EFPs) in the single-point DFT/TD-DFT and CC/EOMCC calculations described in the main text.

| | X | Y | Z | | X | Y | Z | |
|---|---|---|---|---|---|---|---|---|
| C | -0.0537141596 | -2.3670337572 | 1.5390063941 | H | -9.4645466166 | -0.9186451420 | -0.7444525159 | |
| C | 0.6621682126 | -1.4509293066 | 0.6970515529 | H | -9.7610968756 | -0.9320000787 | 1.7332489497 | |
| C | -0.0902745887 | -0.6085586614 | -0.1943833723 | H | -8.0575099000 | -0.5048384610 | 1.9623523375 | |
| C | -1.4542553088 | -0.6888415655 | -0.2085965848 | H | -9.1566372503 | 0.6369720995 | 1.1743745951 | |
| C | -2.1572580806 | -1.6052502815 | 0.6460022858 | C | -7.6980943448 | 0.4461796471 | -1.8611918533 | |
| C | -1.4176573193 | -2.4539910183 | 1.5323495280 | C | -7.7609846568 | 1.9112913542 | -1.4325209700 | |
| H | 0.5209080993 | -3.0053570904 | 2.2068183513 | H | -8.6481799279 | 0.1565580371 | -2.3177921518 | |
| H | 0.4402013230 | 0.0903044600 | -0.8332063504 | H | -6.9411380830 | 0.3027433740 | -2.6361835432 | |
| H | -1.9307280193 | -3.1556278174 | 2.1842241876 | H | -7.9019573066 | 2.5433795531 | -2.3163813564 | |
| C | -2.4492168322 | 0.1033240429 | -1.0538648733 | H | -8.5992672284 | 2.0942573117 | -0.7538820457 | |
| C | -3.7749690070 | -0.4701635449 | -0.5672462024 | H | -6.8421182747 | 2.2284307953 | -0.9296304754 | |
| C | -5.0355314007 | -0.1372624764 | -0.9581033464 | C | 3.5779004811 | 1.1294560429 | -5.4825628933 | ⎤ |
| C | -6.1734273198 | -0.7941419154 | -0.3720657515 | C | 3.4255690357 | 1.7383849113 | -4.0909070663 | ⎥ |
| C | -5.9285285587 | -1.7832796063 | 0.6356212856 | C | 3.0420658222 | 0.6949203205 | -3.0574147642 | ⎥ |
| C | -4.6570680717 | -2.1240477224 | 1.0260453605 | O | 2.8612234230 | 1.3092630435 | -1.7772085935 | ⎥ |
| C | -3.5388656586 | -1.4726824223 | 0.4343217145 | H | 3.8542188364 | 1.8936177389 | -6.2166074738 | ⎥ |
| H | -5.1771077247 | 0.6355775109 | -1.7028625905 | H | 2.6429559077 | 0.6677275866 | -5.8217010411 | ⎥ EFP |
| H | -6.7597463358 | -2.2957249262 | 1.1020584999 | H | 4.3566130318 | 0.3574190935 | -5.4997777190 | ⎥ |
| H | -4.5147615373 | -2.8871771329 | 1.7854437823 | H | 2.6609806422 | 2.5245024188 | -4.1095229546 | ⎥ |
| C | -2.3480462121 | 1.6102541752 | -0.7702995034 | H | 4.3671164064 | 2.2129444922 | -3.7884869206 | ⎥ |
| H | -1.3678781454 | 1.9931668778 | -1.0713077477 | H | 2.1122872168 | 0.1903210414 | -3.3532985119 | ⎥ |
| H | -3.1114225204 | 2.1634483653 | -1.3298166850 | H | 3.8268873447 | -0.0688300598 | -2.9850930796 | ⎥ |
| H | -2.4905638977 | 1.8233215268 | 0.2947075010 | H | 2.7511482323 | 0.5828108851 | -1.0772217457 | ⎦ |
| C | -2.2488662017 | -0.1665305007 | -2.5539920106 | C | 6.2349753847 | 6.3421462002 | -1.3701679874 | ⎤ |
| H | -2.9980675869 | 0.3675042831 | -3.1492029172 | C | 6.2812262946 | 4.8608290844 | -1.0036483948 | ⎥ |
| H | -1.2601713564 | 0.1763208609 | -2.8754979513 | C | 4.9583220115 | 4.1645198367 | -1.2654676463 | ⎥ |
| H | -2.3307657072 | -1.2341332616 | -2.7826977247 | O | 5.0621719474 | 2.7925238931 | -0.8974654766 | ⎥ |
| C | 2.0837986955 | -1.4160568634 | 0.7881149791 | H | 7.1978446429 | 6.8256421370 | -1.1731612271 | ⎥ |
| H | 2.5196794653 | -2.1003375144 | 1.5283461079 | H | 5.4726441136 | 6.8758737100 | -0.7899199062 | ⎥ EFP |
| N | 2.8893618545 | -0.6445733742 | 0.1013323162 | H | 6.0052666333 | 6.4850295191 | -2.4329852810 | ⎥ |
| C | 4.3095261234 | -0.8171086192 | 0.3741338485 | H | 6.5384360719 | 4.7469310580 | 0.0567259016 | ⎥ |
| H | 4.4813708438 | -1.0033250095 | 1.4467799870 | H | 7.0672418069 | 4.3567795792 | -1.5794368343 | ⎥ |
| H | 4.8219465666 | 0.1206474627 | 0.1289753394 | H | 4.1613277015 | 4.6509730456 | -0.6849418988 | ⎥ |
| C | 4.9446569939 | -1.9561754527 | -0.4324587310 | H | 4.6952081176 | 4.2516112811 | -2.3294391641 | ⎥ |
| H | 4.7516064702 | -1.7914521490 | -1.5005980902 | H | 4.2687642519 | 2.3039737538 | -1.2137143557 | ⎦ |
| H | 4.4469723233 | -2.8991860091 | -0.1677202969 | C | 0.1518700042 | 6.6405098409 | -0.8190317438 | ⎤ |
| C | 6.4484434618 | -2.0798490469 | -0.1924005324 | C | -0.1603044906 | 5.2397406088 | -1.3390139478 | ⎥ |
| H | 6.6358238873 | -2.2223208945 | 0.8805019666 | C | 1.0652355264 | 4.3441352583 | -1.3549462514 | ⎥ |
| H | 6.9384059869 | -1.1362781800 | -0.4681224263 | O | 0.6991690161 | 3.0552635909 | -1.8300431511 | ⎥ |
| C | 7.0832325971 | -3.2276782252 | -0.9734230650 | H | -0.7435204163 | 7.2713935322 | -0.8272437462 | ⎥ |
| H | 8.1623610906 | -3.2855005780 | -0.7901184442 | H | 0.9108524650 | 7.1373792129 | -1.4354225823 | ⎥ EFP |
| H | 6.6461779243 | -4.1924629272 | -0.6881690150 | H | 0.5264495048 | 6.6117866364 | 0.2112899358 | ⎥ |
| H | 6.9364890652 | -3.1033678562 | -2.0532770484 | H | -0.5665215644 | 5.3022674857 | -2.3564292148 | ⎥ |
| N | -7.4398730031 | -0.4913443518 | -0.7654545700 | H | -0.9315906929 | 4.7702743832 | -0.7160548359 | ⎥ |
| C | -8.6076770049 | -1.0562946757 | -0.0819751711 | H | 1.8381113446 | 4.7842180334 | -2.0023654777 | ⎥ |
| C | -8.9054051970 | -0.4228882222 | 1.2761662148 | H | 1.4835299403 | 4.2680625112 | -0.3402540898 | ⎥ |
| H | -8.4770298339 | -2.1385578735 | 0.0177864408 | H | 1.4809610489 | 2.4598351991 | -1.7928526986 | ⎦ |

$\boldsymbol{\mu}_0 = -6.425\mathbf{i} - 0.591\mathbf{j} + 0.282\mathbf{k}, \quad \boldsymbol{\mu}_1 = -19.520\mathbf{i} + 0.431\mathbf{j} - 1.386\mathbf{k}$

**FR0-SB in *i*-propanol, CAM-B3LYP/6-31+G*/SMD optimized geometry of the S$_0$ state (in Å) along with the electronic dipole moments (in Debye) characterizing the S$_0$ and S$_1$ electronic states of the FR0-SB chromophore, calculated using the CC/EOMCC-based protocol described in the main text**

The curly brackets indicate the three alcohol solvent molecules being replaced by effective fragment potentials (EFPs) in the single-point DFT/TD-DFT and CC/EOMCC calculations described in the main text.

|   | X | Y | Z |   | X | Y | Z |   |
|---|---|---|---|---|---|---|---|---|
| C |  0.0248039283 | -3.1267966368 |  0.7995283445 | H | -9.1149966649 | -0.1676619283 | -1.3203197781 | |
| C |  0.8011198076 | -1.9832107930 |  0.5662729545 | H | -9.7492531259 | -1.3960695842 |  0.7543012599 | |
| C |  0.1689575049 | -0.7836284490 |  0.1899005940 | H | -8.0981135409 | -1.2430327923 |  1.3768691490 | |
| C | -1.2050100037 | -0.7573303846 |  0.0451838606 | H | -9.0904966080 |  0.2025675933 |  1.1400704946 | |
| C | -1.9724268631 | -1.9189591701 |  0.2679146159 | C | -7.2315496311 |  1.5032212876 | -1.2917595486 | |
| C | -1.3580274433 | -3.1077405071 |  0.6508483160 | C | -7.2608115767 |  2.5162366379 | -0.1478744456 | |
| H |  0.5163985698 | -4.0498813462 |  1.0980159048 | H | -8.1507728705 |  1.5940755265 | -1.8779617110 | |
| H |  0.7662631227 |  0.1080762947 |  0.0244425856 | H | -6.4178537527 |  1.7406213132 | -1.9856752090 | |
| H | -1.9377205231 | -4.0087224044 |  0.8330637791 | H | -7.2810526271 |  3.5356329992 | -0.5508589944 | |
| C | -2.1032048598 |  0.4131890761 | -0.3447149100 | H | -8.1503038635 |  2.3855090922 |  0.4768626935 | |
| C | -3.4774836372 | -0.2485626705 | -0.3315139866 | H | -6.3818266311 |  2.4267414025 |  0.4984169029 | |
| C | -4.6997615059 |  0.3332362492 | -0.6161108863 | C |  3.7195774688 |  1.6439472117 | -1.6127762983 | ⎫ |
| C | -5.8880577299 | -0.4368385459 | -0.5593777045 | C |  2.6145175337 |  1.1127929853 | -2.5123493793 | ⎪ |
| C | -5.7669248580 | -1.7982469595 | -0.1788579533 | C |  4.0150159588 |  3.1105659032 | -1.8660580033 | ⎪ |
| C | -4.5350943439 | -2.3696779446 |  0.1052293537 | O |  3.3635918353 |  1.5003698100 | -0.2220194622 | ⎪ |
| C | -3.3778280143 | -1.5984138472 |  0.0291985458 | H |  4.6356189517 |  1.0630840819 | -1.7859793967 | ⎪ |
| H | -4.7381937397 |  1.3842226776 | -0.8743816530 | H |  2.3944216294 |  0.0613796758 | -2.3003425493 | ⎬ EFP |
| H | -6.6464796055 | -2.4244937579 | -0.1050139767 | H |  1.6949716134 |  1.6933167716 | -2.3797089723 | ⎪ |
| H | -4.4920617987 | -3.4193786702 |  0.3858074052 | H |  2.9127249728 |  1.1853855834 | -3.5642622750 | ⎪ |
| C | -2.0182706599 |  1.5486347870 |  0.6886011339 | H |  4.8067604017 |  3.4734572516 | -1.2034505938 | ⎪ |
| H | -1.0130357307 |  1.9817970381 |  0.7159913526 | H |  4.3499860795 |  3.2540619923 | -2.8988134237 | ⎪ |
| H | -2.7233836873 |  2.3495525062 |  0.4383493320 | H |  3.1194584313 |  3.7218423353 | -1.7110306147 | ⎪ |
| H | -2.2619088888 |  1.1872202623 |  1.6936117314 | H |  3.1669235029 |  0.5436307813 | -0.0274133765 | ⎭ |
| C | -1.7553533953 |  0.9477514262 | -1.7428648448 | C |  5.6506343685 |  2.5582551768 |  2.4979478476 | ⎫ |
| H | -2.4443961736 |  1.7476183318 | -2.0355693768 | C |  4.9146983386 |  1.6277094097 |  3.4538835349 | ⎪ |
| H | -0.7421930175 |  1.3620813335 | -1.7573102615 | C |  7.0539298603 |  2.8843045323 |  2.9769044553 | ⎪ |
| H | -1.8098032130 |  0.1584002966 | -2.4999663040 | O |  5.7765778541 |  1.9693029801 |  1.2026155152 | ⎪ |
| C |  2.2532578822 | -2.1181863495 |  0.7261132752 | H |  5.0846959588 |  3.4974917795 |  2.4068116740 | ⎪ |
| H |  2.5753627058 | -3.0981028442 |  1.1065187108 | H |  3.9225193327 |  1.3653034521 |  3.0689082956 | ⎬ EFP |
| N |  3.1309948817 | -1.2320018794 |  0.4593945257 | H |  5.4784439868 |  0.6990864232 |  3.5995679239 | ⎪ |
| C |  4.5284816135 | -1.6205254775 |  0.6495339655 | H |  4.7761868475 |  2.1031778995 |  4.4321161564 | ⎪ |
| H |  4.6091474974 | -2.4390185453 |  1.3798240281 | H |  7.5554810237 |  3.5575739613 |  2.2736990153 | ⎪ |
| H |  5.0661462353 | -0.7597142584 |  1.0603828350 | H |  7.0262850900 |  3.3726948653 |  3.9570196504 | ⎪ |
| C |  5.1742244406 | -2.0491325689 | -0.6687725704 | H |  7.6517295249 |  1.9696241731 |  3.0644965200 | ⎪ |
| H |  5.0915793958 | -1.2334175922 | -1.3982927791 | H |  4.8918134153 |  1.8433653095 |  0.7984904917 | ⎭ |
| H |  4.6076034020 | -2.8946817782 | -1.0818207256 | C |  1.7636094341 |  4.0069304328 |  1.9095077260 | ⎫ |
| C |  6.6406386657 | -2.4473954366 | -0.5048058410 | C |  2.5913477399 |  5.1044538565 |  1.2524210248 | ⎪ |
| H |  6.7152749993 | -3.2651167125 |  0.2243790200 | C |  0.5371763768 |  4.5622616835 |  2.6111592265 | ⎪ |
| H |  7.2049283623 | -1.6053687963 | -0.0835909048 | O |  1.3042206751 |  3.0557529261 |  0.9474463372 | ⎪ |
| C |  7.2828883119 | -2.8817133229 | -1.8196545596 | H |  2.3869869028 |  3.4871228888 |  2.6529557716 | ⎪ |
| H |  8.3310858582 | -3.1666286754 | -1.6750127704 | H |  3.4631091406 |  4.6905946082 |  0.7353296361 | ⎬ EFP |
| H |  6.7607605198 | -3.7447082240 | -2.2506134444 | H |  1.9895468371 |  5.6512059097 |  0.5171651112 | ⎪ |
| H |  7.2580482117 | -2.0740076882 | -2.5612193773 | H |  2.9558001775 |  5.8192194604 |  1.9996192303 | ⎪ |
| N | -7.1149025861 |  0.1085680392 | -0.8812389775 | H | -0.0365774377 |  3.7606114899 |  3.0868759926 | ⎪ |
| C | -8.3460006876 | -0.6571331383 | -0.7166406737 | H |  0.8260416700 |  5.2814182276 |  3.3855135493 | ⎪ |
| C | -8.8435115610 | -0.7791173402 |  0.7235060844 | H | -0.1165210449 |  5.0726237260 |  1.8941664736 | ⎪ |
| H | -8.2180703868 | -1.6526325807 | -1.1573597291 | H |  2.0656413879 |  2.5902801947 |  0.5407261691 | ⎭ |

$\boldsymbol{\mu}_0 = -4.230\mathbf{i} - 0.367\mathbf{j} - 0.065\mathbf{k}, \quad \boldsymbol{\mu}_1 = -15.328\mathbf{i} + 2.609\mathbf{j} - 1.959\mathbf{k}$

**FR0-SB in *i*-propanol, CAM-B3LYP/6-31+G\*/SMD optimized geometry of the $S_1$ state (in Å) along with the electronic dipole moments (in Debye) characterizing the $S_0$ and $S_1$ electronic states of the FR0-SB chromophore, calculated using the CC/EOMCC-based protocol described in the main text**

The curly brackets indicate the three alcohol solvent molecules being replaced by effective fragment potentials (EFPs) in the single-point DFT/TD-DFT and CC/EOMCC calculations described in the main text.

| | X | Y | Z | | X | Y | Z | |
|---|---|---|---|---|---|---|---|---|
| C | -0.0128720322 | -3.1516474760 | 0.7207268604 | H | -9.1150308906 | -0.1968252205 | -1.3150656408 | |
| C | 0.7967782840 | -1.9881570263 | 0.4977245634 | H | -9.7356558960 | -1.4535401077 | 0.7393679587 | |
| C | 0.1438654385 | -0.7584476288 | 0.1344880245 | H | -8.0815978259 | -1.3179454387 | 1.3599705163 | |
| C | -1.2154931782 | -0.7300837207 | 0.0005988451 | H | -9.0724655058 | 0.1346725412 | 1.1562182877 | |
| C | -2.0123716555 | -1.9077504060 | 0.2127609394 | C | -7.2501713395 | 1.4817438058 | -1.2852876598 | |
| C | -1.3730119628 | -3.1346456339 | 0.5872995072 | C | -7.2871937049 | 2.4835589125 | -0.1325373236 | |
| H | 0.4857491003 | -4.0766674445 | 1.0029009493 | H | -8.1737148573 | 1.5552600674 | -1.8644217255 | |
| H | 0.7451798390 | 0.1313915918 | -0.0210122912 | H | -6.4424777861 | 1.7256195601 | -1.9810179130 | |
| H | -1.9566638455 | -4.0345324654 | 0.7614181847 | H | -7.3303446607 | 3.5008203560 | -0.5371782282 | |
| C | -2.1154504128 | 0.4473113875 | -0.3692633536 | H | -8.1678136311 | 2.3365186732 | 0.4997705996 | |
| C | -3.4890154617 | -0.2145364679 | -0.3627446754 | H | -6.3997238024 | 2.4089244949 | 0.5032162181 | |
| C | -4.6992617925 | 0.3448171283 | -0.6353175823 | C | 3.7099283798 | 1.6374188673 | -1.5512387180 | ⎤ |
| C | -5.8961533570 | -0.4523911318 | -0.5759821743 | C | 2.6009465861 | 1.1285832962 | -2.4595648624 | ⎥ |
| C | -5.7625320074 | -1.8313846816 | -0.2096507721 | C | 4.0356920922 | 3.0974297083 | -1.8102290227 | ⎥ |
| C | -4.5425208613 | -2.3968164547 | 0.0670151118 | O | 3.3463988380 | 1.5001779061 | -0.1649722190 | ⎥ |
| C | -3.3640954305 | -1.6012090831 | -0.0059167548 | H | 4.6147637265 | 1.0380189511 | -1.7225905931 | ⎥ |
| H | -4.7572431160 | 1.3957137564 | -0.8883700025 | H | 2.3575686560 | 0.0836889892 | -2.2440492461 | ⎥ EFP |
| H | -6.6408338277 | -2.4607788189 | -0.1508125762 | H | 1.6928067128 | 1.7286489099 | -2.3345633075 | ⎥ |
| H | -4.4832364101 | -3.4474542776 | 0.3351311656 | H | 2.9089943960 | 1.1934438116 | -3.5094493724 | ⎥ |
| C | -2.0309422689 | 1.5655860171 | 0.6824285685 | H | 4.8307072685 | 3.4481723311 | -1.1451326045 | ⎥ |
| H | -1.0213116189 | 1.9877081815 | 0.7204886501 | H | 4.3775799823 | 3.2302217245 | -2.8423442619 | ⎥ |
| H | -2.7280151166 | 2.3767416979 | 0.4418587254 | H | 3.1511022881 | 3.7263195456 | -1.6618783697 | ⎥ |
| H | -2.2814317391 | 1.1897020986 | 1.6803018331 | H | 3.1461288324 | 0.5329930141 | 0.0352422996 | ⎦ |
| C | -1.7734486890 | 1.0037278082 | -1.7598939144 | C | 5.7005455573 | 2.6022178037 | 2.4803867988 | ⎤ |
| H | -2.4563260805 | 1.8148436605 | -2.0364915883 | C | 5.0182024434 | 1.6840937685 | 3.4863730116 | ⎥ |
| H | -0.7561911862 | 1.4072465225 | -1.7692443633 | C | 7.1151640360 | 2.9619686569 | 2.8980271428 | ⎥ |
| H | -1.8367327893 | 0.2275919816 | -2.5296058795 | O | 5.7880292370 | 1.9852513651 | 1.1960380434 | ⎥ |
| C | 2.2096518540 | -2.1129749876 | 0.6495007021 | H | 5.1155945838 | 3.5297284490 | 2.3897758452 | ⎥ |
| H | 2.5538823214 | -3.1105299303 | 0.9560172312 | H | 4.0160889146 | 1.3985538955 | 3.1472194782 | ⎥ EFP |
| N | 3.1028759983 | -1.1756501995 | 0.4675917936 | H | 5.6014445330 | 0.7670882041 | 3.6290588822 | ⎥ |
| C | 4.4876222221 | -1.5871881269 | 0.6490022328 | H | 4.9126771115 | 2.1805394893 | 4.4582693989 | ⎥ |
| H | 4.5663146345 | -2.3786905861 | 1.4121375452 | H | 7.5780491056 | 3.6295213100 | 2.1634109464 | ⎥ |
| H | 5.0575459657 | -0.7272587551 | 1.0198780430 | H | 7.1172885580 | 3.4689696963 | 3.8690037099 | ⎥ |
| C | 5.1295221433 | -2.0886951329 | -0.6504773065 | H | 7.7320169573 | 2.0593336730 | 2.9795593743 | ⎥ |
| H | 5.0554735989 | -1.3027790610 | -1.4134202566 | H | 4.8908295067 | 1.8383257227 | 0.8246122244 | ⎦ |
| H | 4.5493153911 | -2.9427636564 | -1.0263027215 | C | 1.7856602390 | 4.0259553190 | 1.9340098220 | ⎤ |
| C | 6.5907839027 | -2.4995354565 | -0.4728633849 | C | 2.6235620055 | 5.1123204285 | 1.2707839665 | ⎥ |
| H | 6.6551998142 | -3.2982199862 | 0.2785368448 | C | 0.5728261437 | 4.6009937983 | 2.6438843335 | ⎥ |
| H | 7.1624677135 | -1.6523386949 | -0.0713476930 | O | 1.3069918022 | 3.0806872017 | 0.9769699955 | ⎥ |
| C | 7.2355166845 | -2.9748780762 | -1.7725313444 | H | 2.4074360484 | 3.4991782987 | 2.6739899567 | ⎥ |
| H | 8.2792463531 | -3.2707248070 | -1.6153006402 | H | 3.4864105925 | 4.6863003284 | 0.7493509817 | ⎥ EFP |
| H | 6.7049018061 | -3.8411151555 | -2.1865930848 | H | 2.0239236364 | 5.6651062167 | 0.5381738694 | ⎥ |
| H | 7.2263434371 | -2.1852685522 | -2.5337968991 | H | 3.0007004106 | 5.8239746413 | 2.0148113998 | ⎥ |
| N | -7.1122225734 | 0.0832653330 | -0.8693671952 | H | -0.0069574199 | 3.8105398725 | 3.1310020390 | ⎥ |
| C | -8.3457088201 | -0.6946668083 | -0.7219803210 | H | 0.8780088405 | 5.3222717918 | 3.4099557841 | ⎥ |
| C | -8.8289453623 | -0.8387515442 | 0.7204978354 | H | -0.0809600811 | 5.1142297527 | 1.9289697673 | ⎥ |
| H | -8.2062575402 | -1.6783561006 | -1.1816624778 | H | 2.0588679629 | 2.5920711074 | 0.5758653536 | ⎦ |

$\boldsymbol{\mu}_0 = -6.357\mathbf{i} - 0.160\mathbf{j} - 0.505\mathbf{k}, \quad \boldsymbol{\mu}_1 = -18.823\mathbf{i} + 2.477\mathbf{j} - 2.284\mathbf{k}$

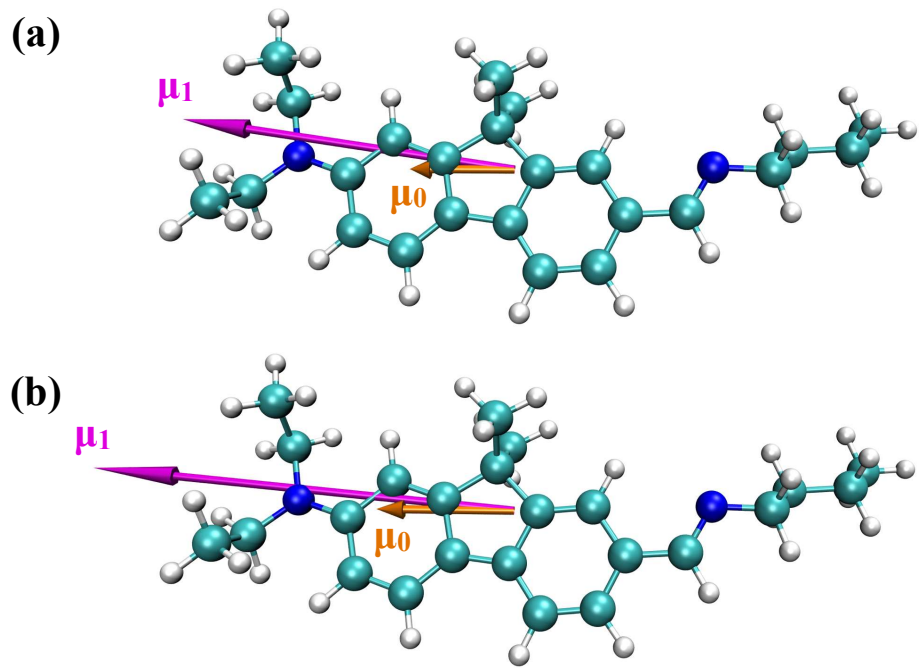

**Fig. S7**. The CAM-B3LYP/6-31+G* optimized structures of **FR0**-SB in its (a) $S_0$ and (b) $S_1$ states. For each structure, the $\boldsymbol{\mu_0}$ (shorter orange vector) and $\boldsymbol{\mu_1}$ (longer magenta vector) dipole moments characterizing the $S_0$ and $S_1$ states, respectively, of the **FR0**-SB chromophore, calculated using the CC/EOMCC-based protocol described in the main text, are depicted as well.

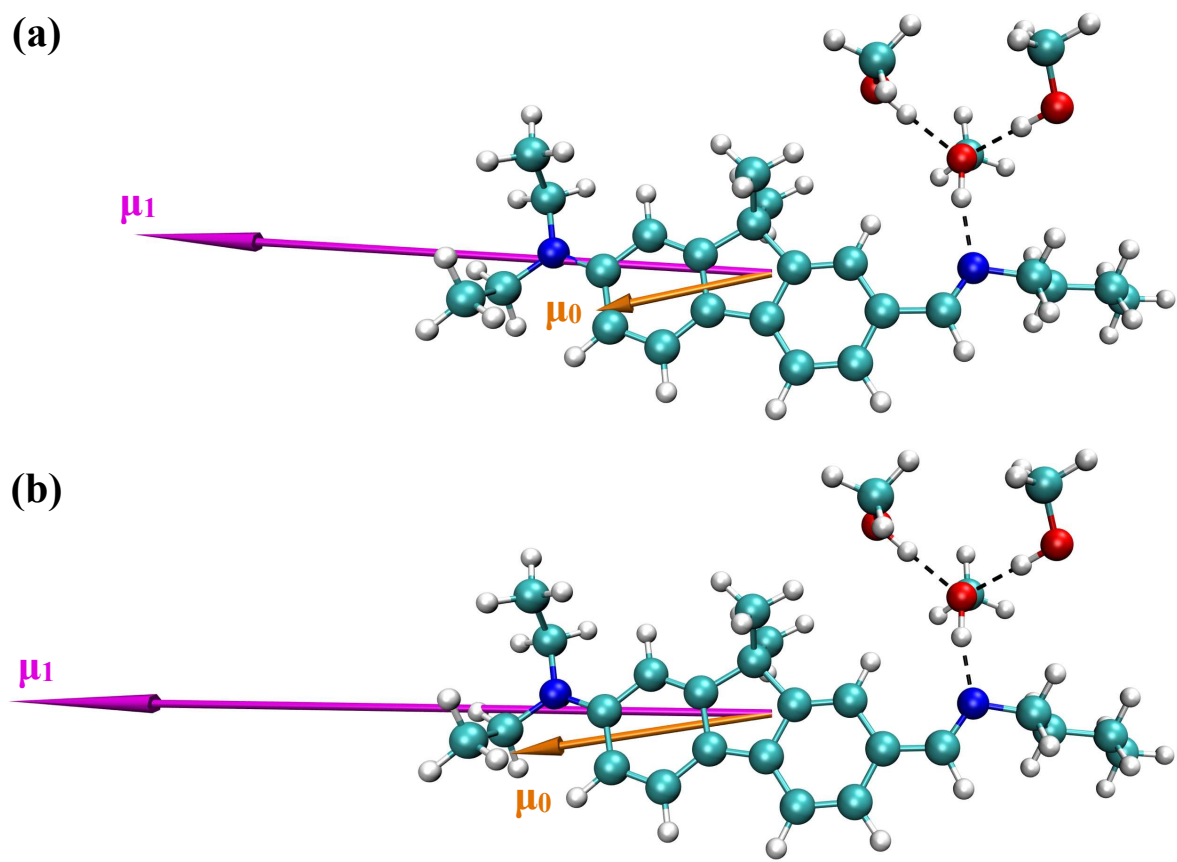

**Fig. S8**. The CAM-B3LYP/6-31+G*/SMD optimized structures of **FR0**-SB in its (a) $S_0$ and (b) $S_1$ states hydrogen-bonded to a cluster of three methanol solvent molecules. For each structure, the $\mu_0$ (shorter orange vector) and $\mu_1$ (longer magenta vector) dipole moments characterizing the $S_0$ and $S_1$ states, respectively, of the **FR0**-SB chromophore, calculated using the CC/EOMCC-based protocol described in the main text, are depicted as well.

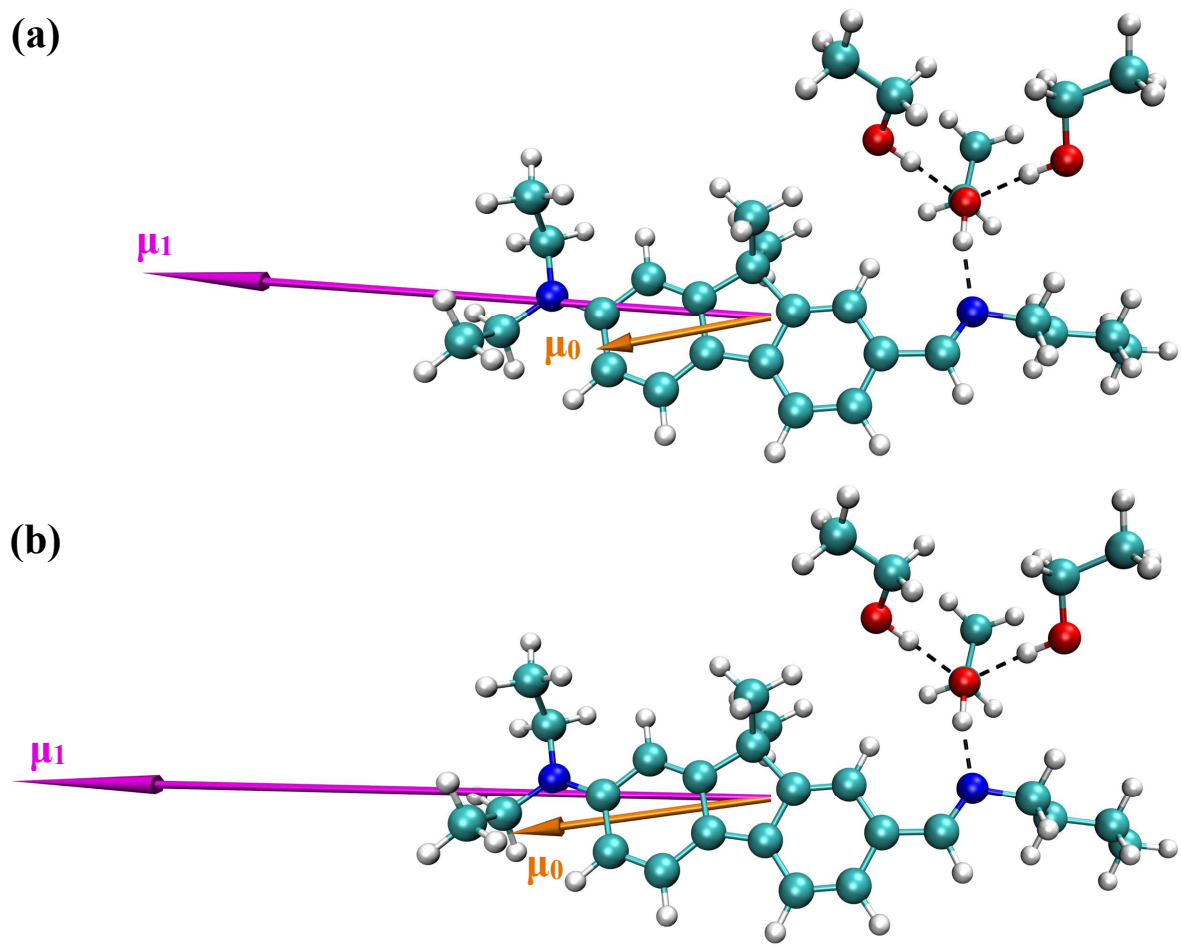

**Fig. S9**. The CAM-B3LYP/6-31+G*/SMD optimized structures of **FR0**-SB in its (a) $S_0$ and (b) $S_1$ states hydrogen-bonded to a cluster of three ethanol solvent molecules. For each structure, the **$\mu_0$** (shorter orange vector) and **$\mu_1$** (longer magenta vector) dipole moments characterizing the $S_0$ and $S_1$ states, respectively, of the **FR0**-SB chromophore, calculated using the CC/EOMCC-based protocol described in the main text, are depicted as well.

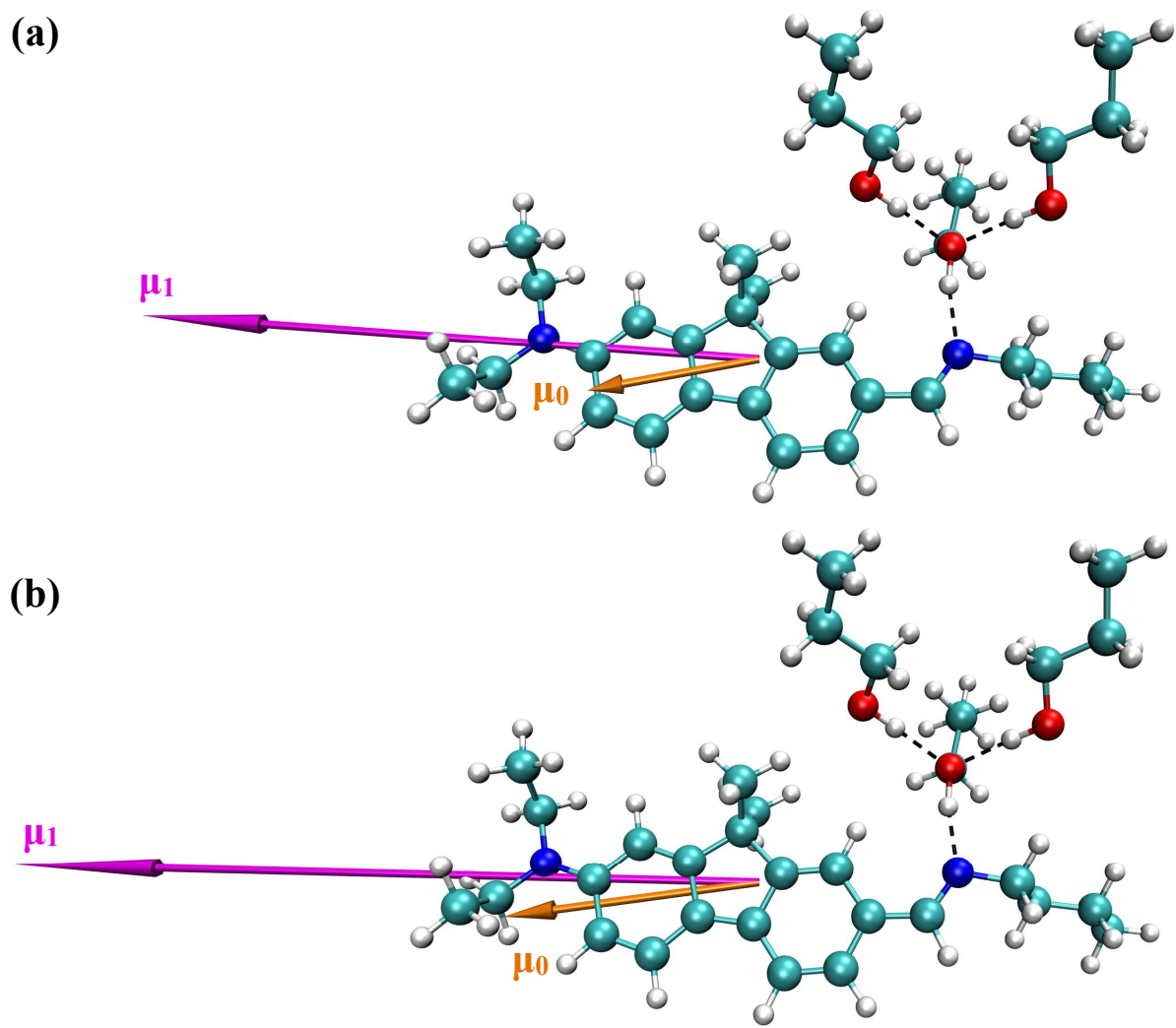

**Fig. S10**. The CAM-B3LYP/6-31+G*/SMD optimized structures of **FR0**-SB in its (a) $S_0$ and (b) $S_1$ states hydrogen-bonded to a cluster of three *n*-propanol solvent molecules. For each structure, the **μ₀** (shorter orange vector) and **μ₁** (longer magenta vector) dipole moments characterizing the $S_0$ and $S_1$ states, respectively, of the **FR0**-SB chromophore, calculated using the CC/EOMCC-based protocol described in the main text, are depicted as well.

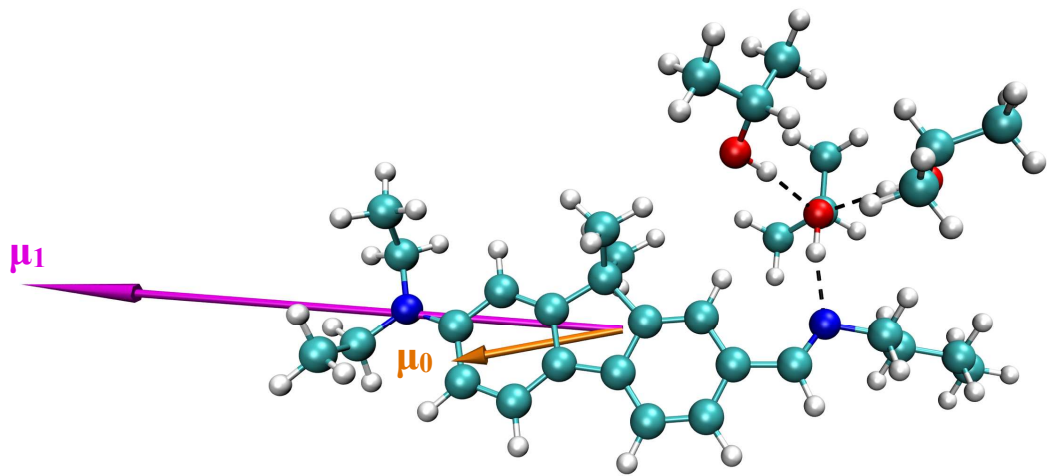

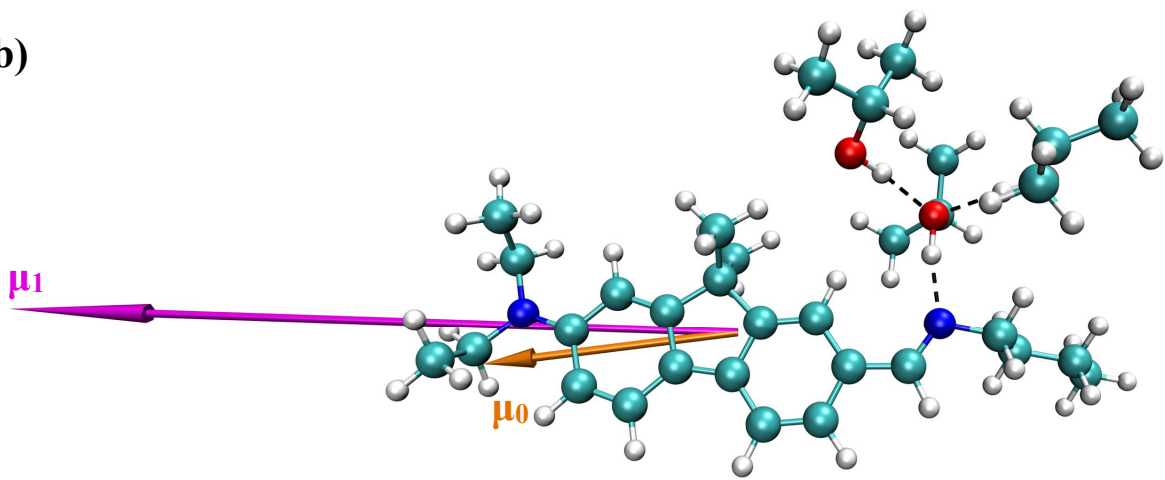

**Fig. S11**. The CAM-B3LYP/6-31+G*/SMD optimized structures of **FR0**-SB in its (a) $S_0$ and (b) $S_1$ states hydrogen-bonded to a cluster of three *i*-propanol solvent molecules. For each structure, the **μ₀** (shorter orange vector) and **μ₁** (longer magenta vector) dipole moments characterizing the $S_0$ and $S_1$ states, respectively, of the **FR0**-SB chromophore, calculated using the CC/EOMCC-based protocol described in the main text, are depicted as well.

# The Dominant Orbitals Defining the $S_0$–$S_1$ Transitions in the [FR0-SB⋯HOR] Complexes

In this section, we are showing the highest-occupied and lowest-unoccupied molecular orbitals, HOMO and LUMO, respectively, that dominate the $S_0$–$S_1$ transitions in the [**FR0**-SB⋯HOR] complexes, using **FR0**-SB hydrogen-bonded to a cluster of three methanol solvent molecules as a representative example (see Fig. S12). The HOMO and LUMO of the [**FR0**-SB⋯HOMe] system, obtained in the Kohn–Sham CAM-B3LYP/6-31+G*/SMD calculations at the $S_0$ minimum-energy structure, indicate that there is no charge transfer between the **FR0**-SB chromophore and the methanol solvent molecules upon the $S_0 \rightarrow S_1$ photoexcitation.

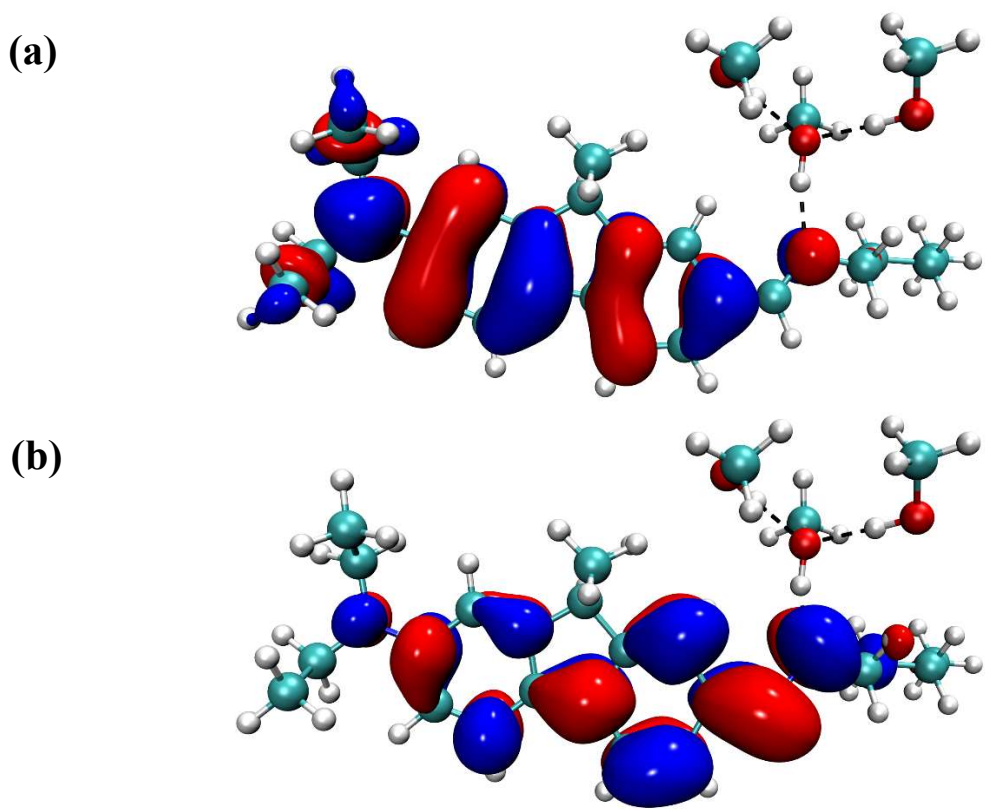

**(a)**

**(b)**

**Fig. S12**. The (a) HOMO and (b) LUMO that characterize the $S_0 \rightarrow S_1$ photoexcitation in the complex of **FR0**-SB hydrogen-bonded to a cluster of three methanol solvent molecules. The one-electron HOMO to LUMO excitation having an amplitude of 0.95 dominates the $S_0 \rightarrow S_1$ transition. The HOMO and LUMO correspond to $\pi$ and $\pi^*$ molecular orbitals, respectively, localized on the **FR0**-SB chromophore. In plotting these orbitals, we chose an isovalue of 0.02.